\documentclass[journal, twoside]{IEEEtran}
\usepackage{graphicx}
\usepackage{graphics,epsf}
\usepackage{amsmath}
\usepackage{bm}
\usepackage{amssymb}
\usepackage{float}
\newcommand{\bX}{\ensuremath{\mathbf{X}}}
\newcommand{\bY}{\ensuremath{\mathbf{Y}}}
\newcommand{\bZ}{\ensuremath{\mathbf{Z}}}
\newcommand{\bH}{\ensuremath{\mathbf{H}}}

\newcommand{\snr}{\ensuremath{\mathrm{SNR}}}
\newcommand{\dmt}{DMT}
\newcommand{\mimo}{MIMO}
\newcommand{\simo}{SIMO}
\newcommand{\miso}{MISO}
\newcommand{\awgn}{AWGN}
\newcommand{\af}{AF}
\newcommand{\df}{DF}
\newcommand{\naf}{NAF}
\newcommand{\maf}{MAF}
\newcommand{\ddf}{DDF}
\newcommand{\cf}{CF}
\newcommand{\marc}{MARC}
\newcommand{\csi}{CSI}
\newtheorem{theorem}{Theorem}
\newtheorem{corollary}{Corollary}
\newtheorem{proposition}{Proposition}
\newtheorem{lemma}{Lemma}
\newtheorem{conjecture}{Conjecture}

\begin{document}
\title{Multi-Antenna Cooperative Wireless Systems: \\A Diversity-Multiplexing Tradeoff Perspective}
\author{Melda~Yuksel,~\IEEEmembership{Student Member,~IEEE,}
        and~Elza~Erkip,~\IEEEmembership{Senior~Member,~IEEE}
\thanks{Manuscript received August 31, 2006.; revised February 16, 2007. This material is based upon work partially
supported by the National Science Foundation under Grant No.
0093163. The material in this paper was presented in part at the
40th Annual Conference on Information Sciences and Systems, CISS
2006, IEEE International Symposium on Information Theory, Seattle,
WA, July 2006, and IEEE International Conference on Communications,
Glasgow, June 2007.}
\thanks{ The authors are with Electrical and Computer Engineering
Department, Polytechnic University, Brooklyn, NY 11201 USA (e-mail:
myukse01@utopia.poly.edu; elza@poly.edu).}}

\markboth{IEEE TRANSACTIONS ON INFORMATION THEORY,~Vol.~x,
No.~x,~DECEMBER~2007}{Yuksel and Erkip: Multi-Antenna Cooperative
Wireless Systems: A Diversity-Multiplexing Tradeoff Perspective}
\maketitle

\begin{abstract}
We consider a general multiple antenna network with multiple
sources, multiple destinations and multiple relays in terms of the
diversity-multiplexing tradeoff (\dmt). We examine several subcases
of this most general problem taking into account the processing
capability of the relays (half-duplex or full-duplex), and the
network geometry (clustered or non-clustered). We first study the
multiple antenna relay channel with a full-duplex relay to
understand the effect of increased degrees of freedom in the direct
link. We find \dmt\ upper bounds and investigate the achievable
performance of decode-and-forward (\df), and compress-and-forward
(\cf) protocols. Our results suggest that while \df\ is \dmt\
optimal when all terminals have one antenna each, it may not
maintain its good performance when the degrees of freedom in the
direct link is increased, whereas \cf\ continues to perform
optimally. We also study the multiple antenna relay channel with a
half-duplex relay. We show that the half-duplex \dmt\ behavior can
significantly be different from the full-duplex case. We find that
\cf\ is \dmt\ optimal for half-duplex relaying as well, and is the
first protocol known to achieve the half-duplex relay \dmt. We next
study the multiple-access relay channel (\marc) \dmt. Finally, we
investigate a system with a single source-destination pair and
multiple relays, each node with a single antenna, and show that even
under the idealistic assumption of full-duplex relays and a
clustered network, this virtual multi-input multi-output (\mimo)
system can never fully mimic a real \mimo\ \dmt. For cooperative
systems with multiple sources and multiple destinations the same
limitation remains to be in effect.
\end{abstract}

\begin{keywords}
cooperation, diversity-multiplexing tradeoff, fading channels,
multiple-input multiple-output (\mimo), relay channel, wireless
networks.
\end{keywords}
\IEEEpeerreviewmaketitle

\section{Introduction}\label{sec:Introduction}
Next-generation wireless communication systems demand both high
transmission rates and a quality-of-service guarantee. This demand
directly conflicts with the properties of the wireless medium. As a
result of the scatterers in the environment and mobile terminals,
signal components received over different propagation paths may add
destructively or constructively and cause random fluctuations in the
received signal strength~\cite{Rappaport}. This phenomena, which is
called fading, degrades the system performance. Multi-input
multi-output (\mimo) systems introduce spatial diversity to combat
fading. Additionally, taking advantage of the rich scattering
environment, \mimo\ increases spatial
multiplexing~\cite{Foschini96,Telatar99}.

User cooperation/relaying is a practical alternative to \mimo\ when
the size of the wireless device is limited. Similar to \mimo,
cooperation among different users can increase the achievable rates
and decrease susceptibility to channel
variations~\cite{SendonarisEA03_1,SendonarisEA03_2}. In
\cite{LanemanTW02}, the authors proposed relaying strategies that
increase the system reliability. Although the capacity of the
general relay channel problem has been unsolved for over thirty
years~\cite{VanDerMeulen71,CoverElG79}, the papers
\cite{SendonarisEA03_1,SendonarisEA03_2} and \cite{LanemanTW02}
triggered a vast literature on cooperative wireless systems. Various
relaying strategies and space-time code designs that increase
diversity gains or achievable rates are studied
in~\cite{BoyerFY04}\nocite{GastparV02,GuptaK03,HasnaA022,HostMadsen06,HostMadsenZ05,JananiHHN04,
JindalMG04,KatzShamai05,KatzShamai06,KhojastepourSA03,KramerGG04,LaiLiuEG05,LaiLiuEG06,LanemanW03,
LiangK06,LiangV05,MaricYK05,NabarBolcskeiK04,NgGoldsmith04,NgGoldsmith05,NgLanemanGoldsmith06,
ReznikKV04,ScheinG00,SheaWAL03_1,YukselE03,YukselE04_2}-\cite{ZahediMElG04}.

As opposed to the either/or approach of higher reliability or higher
rate, the seminal paper \cite{ZhengT03} establishes the fundamental
tradeoff between these two measures, reliability and rate, also
known as the diversity-multiplexing tradeoff (\dmt), for \mimo\
systems. At high \snr, the measure of reliability is the diversity
gain, which shows how fast the probability of error decreases with
increasing \snr. The multiplexing gain, on the other hand, describes
how fast the actual rate of the system increases with \snr. \dmt\ is
a powerful tool to evaluate the performance of different multiple
antenna schemes at high \snr; it is also a useful performance
measure for cooperative/relay systems. On one hand it is easy enough
to tackle, and on the other hand it is strong enough to show
insightful comparisons among different relaying schemes. While the
capacity of the relay channel is not known in general, it is
possible to find relaying schemes that exhibit optimal \dmt\
performance. Therefore, in this work we study cooperative/relaying
systems from a \dmt\ perspective.

In a general cooperative/relaying network with multiple antenna
nodes, some of the nodes are sources, some are destinations, and
some are mere relays. Finding a complete \dmt\ characterization of
the most general network seems elusive at this time, we will
highlight some of the challenges in the paper. Therefore, we examine
the following important subproblems of the most general network.
\begin{itemize}
    \item \textit{Problem 1:} A single source-destination system, with one
relay, each node has multiple antennas,
    \item \textit{Problem 2:} The multiple-access relay channel with
multiple sources, one destination and one relay, each node has
multiple antennas,
    \item \textit{Problem 3:} A single source-destination system with multiple relays, each node
has a single antenna,
    \item \textit{Problem 4:} A multiple source-multiple destination system, each node has a
single antenna.
\end{itemize}

An important constraint is the processing capability of the
relay(s). We investigate cooperative/relaying systems and strategies
under the full-duplex assumption, i.e. when wireless devices
transmit and receive simultaneously, to highlight some of the
fundamental properties and limitations. Half-duplex systems, where
wireless devices cannot transmit and receive at the same time, are
also of interest, as the half-duplex assumption more accurately
models a practical system. Therefore, we study both full-duplex and
half-duplex relays in the above network configurations.

The channel model and relative node locations have an important
effect on the \dmt\ results that we provide in this paper.
In~\cite{YukselE04}, we investigated \textit{Problem 3} from the
diversity perspective only. We showed that in order to have maximal
\mimo\ diversity gain, the relays should be clustered around the
source and the destination evenly. In other words, half of the
relays should be in close proximity to the source and the rest close
to the destination so that they have a strong inter-user channel
approximated as an additive white Gaussian noise (\awgn) channel.
Only for this clustered case we can get maximal \mimo\ diversity,
any other placement of relays results in lower diversity gains.
Motivated by this fact, we will also study the effect of clustering
on the relaying systems listed above.

\subsection{Related Work}\label{subsec:RelatedWork}
Most of the literature on cooperative communications consider single
antenna terminals. The \dmt\ of relay systems were first studied in
\cite{LanemanTW02} and \cite{AzarianEGS05} for half-duplex relays.
Amplify-and-forward (\af) and decode-and-forward (\df) are two of
the protocols suggested in~\cite{LanemanTW02} for a single relay
system with single antenna nodes. In both protocols, the relay
listens to the source during the first half of the frame, and
transmits during the second half, while the source remains silent.
To overcome the losses of strict time division between the source
and the relay, \cite{LanemanTW02} offers incremental relaying, in
which there is a 1-bit feedback from the destination to both the
source and the relay, and the relay is used only when needed, i.e.
only if the destination cannot decode the source during the first
half of the frame. In \cite{NabarBolcskeiK04}, the authors do not
assume feedback, but to improve the \af\ and \df\ schemes of
\cite{LanemanTW02} they allow the source to transmit simultaneously
with the relay. This idea is also used in \cite{AzarianEGS05} to
study the non-orthogonal amplify-and-forward (\naf) protocol in
terms of \dmt. Later on, a slotted \af\ scheme is proposed
in~\cite{YangBelfiore06}, which outperforms the \naf\ scheme of
\cite{AzarianEGS05} in terms of \dmt. Azarian et al. also propose
the dynamic decode-and-forward (\ddf) protocol
in~\cite{AzarianEGS05}. In \ddf\ the relay listens to the source
until it is able to decode reliably. When this happens, the relay
re-encodes the source message and sends it in the remaining portion
of the frame. The authors find that \ddf\ is optimal for low
multiplexing gains but it is suboptimal when the multiplexing gain
is large. This is because at high multiplexing gains, the relay
needs to listen to the source longer and does not have enough time
left to transmit the high rate source information. This is not an
issue when the multiplexing gain is small as the relay usually
understands the source message at an earlier time instant and has
enough time to transmit.

\mimo\ relay channels are studied in terms of ergodic capacity
in~\cite{WangZHostMadsen05} and in terms of \dmt\
in~\cite{YangBelfiore07}. The latter considers the \naf\ protocol
only, presents a lower bound on the \dmt\ performance and designs
space-time block codes. This lower bound is not tight in general and
is valid only if the number of relay antennas is less than or equal
to the number of source antennas.

The multiple-access relay channel (\marc) is introduced
in~\cite{KramervWijngaarden00,KramerGG04,SankaranarayananKM04}. In
\marc, the relay helps multiple sources simultaneously to reach a
common destination. The \dmt\ for the half-duplex \marc\ with single
antenna nodes is studied in~\cite{AzarianEGS06,ChenL06,ChenAL07}. In
\cite{AzarianEGS06}, the authors find that \ddf\ is \dmt\ optimal
for low multiplexing gains; however, this protocol remains to be
suboptimal for high multiplexing gains analogous to the
single-source relay channel. This region, where \ddf\ is suboptimal,
is achieved by the multiple access amplify and forward (\maf)
protocol \cite{ChenL06,ChenAL07}.

When multiple single antenna relays are present, the
papers~\cite{BoyerFY04,HasnaA022,JananiHHN04, LanemanW03,
NabarBolcskeiK04,SheaWAL03_1,YukselE03} show that diversity gains
similar to multi-input single-output (\miso) or single-input
multi-output (\simo) systems are achievable for Rayleigh fading
channels. Similarly,
\cite{LanemanTW02,AzarianEGS05,PrasadV04,BletsasKRL06} upper bound
the system behavior by \miso\ or \simo\ \dmt\ if all links have
Rayleigh fading. In other words, relay systems behave similar to
either transmit or receive antenna arrays. \textit{Problem 4} is
first analyzed in~\cite{HostMadsenN05} in terms of achievable rates
only, where the authors compare a two-source two-destination
cooperative system with a $2\times 2$ \mimo\ and show that the
former is multiplexing gain limited by 1, whereas the latter has
maximum multiplexing gain of 2.

\subsection{Contributions}\label{subsec:Contributions}
In the light of the related work described in
Section~\ref{subsec:RelatedWork}, we can summarize our contributions
as follows:
\begin{itemize}

\item We study \textit{Problem 1} with full-duplex relays and compare \df\ and
compress-and-forward (\cf) \cite{CoverElG79,KramerGG04} strategies
in terms of \dmt\ for both clustered and non-clustered systems. We
find that there is a fundamental difference between these two
schemes. The \cf\ strategy is \dmt\ optimal for any number of
antennas at the source, the destination or the relay, whereas \df\
is not.

\item We also study \textit{Problem 1} with half-duplex relays. This
study reveals that for half-duplex systems we can find tighter upper
bounds than the full-duplex \dmt\ upper bounds. Moreover, we show
that the \cf\ protocol achieves this half-duplex \dmt\ bound for any
number of antennas at the nodes. This is the first known result on
\dmt\ achieving half-duplex relaying protocols.

\item For \textit{Problem 2} we show that the \cf\ protocol achieves a significant portion of the
half-duplex \dmt\ upper bound for high multiplexing gains. Our
results for single antenna \marc\ easily extend to multiple antenna
terminals.

\item We examine \textit{Problem 3} and \textit{Problem 4} and develop the \dmt\ analysis to understand
if the \textit{network} provides any \mimo\ benefits. Our analysis
shows that even for clustered systems with full-duplex relays, all
relay systems fall short of \mimo, mainly due to multiplexing gain
limitations. The same problem persists in cooperative systems with
multiple source destination pairs.

\end{itemize}

Overall, our work sheds light onto high \snr\ behavior of
cooperative networks as described by the \dmt, and suggests optimal
transmission and relaying strategies.

The paper is organized as follows. Section~\ref{sec:SystemModel}
describes the general system model. In
Section~\ref{sec:preliminaries}, we give some preliminary
information that will be used frequently in the rest of the paper.
In Section~\ref{sec:MultiAntennaSingleRelay} we solve the single
user, single relay problem with multiple antennas for full-duplex
relays, and in Section~\ref{sec:MultiAntennaSingleRelayHD} we solve
the same problem for half-duplex relays (\textit{Problem 1}).
Section~\ref{sec:MARC} introduces \marc, and suggests an achievable
\dmt\ (\textit{Problem 2}). In Section~\ref{sec:SingleAntenna} we
study two problems: the two relay system with a single source
destination pair (\textit{Problem 3}), and the two source two
destination problem (\textit{Problem 4}). Finally, in
Section~\ref{sec:Conclusion} we conclude.

\section{General System Model}\label{sec:SystemModel}
For the most general model all the channels in the system have
independent, slow, frequency non-selective, Rician fading. For
Rician fading channels, the channel gain matrix is written as
\[ \bH = \sqrt{\frac{K}{K+1}}{\bH}_l+\sqrt{\frac{1}{K+1}}{\bH}_s,\]
where $K \geq 0$, ${\bH}_l$ and ${\bH}_s$ denote the Rician factor,
the line of sight component and the scattered component
respectively. The \dmt\ for Rician channels are studied in detail in
\cite{ZhaoMMW06}. In \cite{ZhaoMMW06} the authors find that for
finite Rician factor $K$, the channel mean does not affect the \dmt\
behavior, and the system \dmt\ will be equal to that of a Rayleigh
fading channel with $K=0$. On the other hand in \cite{ShinCL06} the
authors study the effect of $K$ on \miso\ and \simo\ \dmt\ when $K$
approaches infinity. They find that for large $K$, the system
diversity increases linearly with $K$. Moreover, when $K$ tends to
infinity, the diversity gain is infinity for all multiplexing gains
up to $rank(\bH_l)$.

\begin{figure}[t]
\centering
\includegraphics[width=3.4 in]{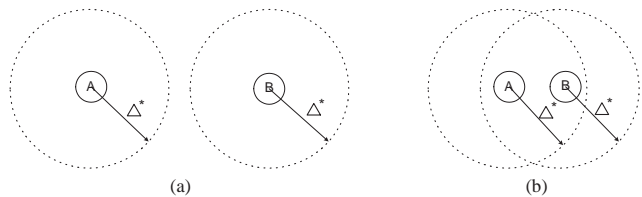}
\caption{Two nodes A and B are in (a) Rayleigh zones, (b) AWGN
zones.} \label{fig:zones}
\end{figure}

Based on the above observations, without loss of generality, in this
work we assume a discrete approximation to the Rician model: If two
nodes are apart more than a threshold distance $\Delta^*$, the line
of sight component is too weak and the Rician factor $K$ can be
assumed to be equal to zero. Thus the channel gain matrix is
distributed as Rayleigh, and we say that the nodes are in
\textit{Rayleigh zones}, Fig.~\ref{fig:zones}(a). On the other hand,
if the inter-node distance is less than $\Delta^*$, the line of
sight component in the received signal is strong; $K$ can be assumed
to be infinity and the Rician distribution approximates a Gaussian.
In this case we say that the nodes are in \textit{\awgn\ zones},
Fig.~\ref{fig:zones}(b).

For the Rayleigh zone, the channel gain matrix for \mimo\ terminals
has independent, identically distributed (i.i.d.) zero mean complex
Gaussian entries with real and imaginary parts each with variance
$\sigma^2$. The variance $\sigma^2$ is proportional to
$1/\Delta^\alpha$, where $\Delta$ denotes the internode distance,
and $\alpha$ is the path loss exponent. If nodes $i$ and $j$ are in
the \awgn\ zone, the channel gain matrix from node $i$ to $j$ has
deterministic entries, all equal to $\sqrt{G_{ij}}$ and the channel
gain matrix has rank 1. There is also a dead zone around the nodes,
which limits the channel gain.

Depending on the locations of the nodes, the Rayleigh or \awgn\ zone
assumption results in two important configurations we will consider:
clustered and non-clustered. For the clustered system, all the
source(s) and some of the relay(s) are in the same \awgn\ zone, and
the destination(s) and the remaining relay(s) are in another \awgn\
zone, but the source cluster and the destination cluster, which are
more than the threshold distance $\Delta^*$ apart, are in their
Rayleigh zones. However, for the non-clustered system, every pair of
nodes in the system are in their Rayleigh zones\footnote{Note that
all mutual information expressions in the paper will be considered
as random quantities. However, if two nodes are clustered, the
channel gains in between these two nodes' antennas take certain
values with probability one.}. We do not explicitly study the
systems in which some nodes are clustered and some are not in this
paper, although our results can easily be applied to these cases as
well.

The relay(s) can be full-duplex, that is they can transmit and
receive at the same time in the same band
(Sections~\ref{sec:MultiAntennaSingleRelay}, and
\ref{sec:SingleAntenna}), or half-duplex
(Sections~\ref{sec:MultiAntennaSingleRelayHD} and \ref{sec:MARC}).
The transmitters (source(s) and relay(s)) in the systems under
consideration have individual power constraints $P_i$. All the noise
vectors at the receivers (relay(s) and destination(s)) have i.i.d.
complex Gaussian entries with zero mean and variance 1. Without loss
of generality we assume the transmit power levels are such that the
average received signal powers at the destination(s) are similar,
and we define \snr\ as the common average received signal to noise
ratio (except for constant multiplicative factors) at the
destination. Because of this assumption, for the clustered systems
we study in Section~\ref{sec:SingleAntenna}, the nodes in the source
cluster hear the transmitters in their cluster much stronger than
the transmitters in the destination cluster, and for all practical
purposes we can ignore the links from the destination cluster to the
nodes in the source cluster. This assumption is the same as the
level set approach of~\cite{GuptaK03}. For non-clustered systems
each node can hear all others.

All the receivers have channel state information (\csi) about their
incoming fading levels\footnote{Because of this assumption, all
mutual information expressions in the paper should be interpreted as
conditioned on the receiver side \csi. We omit this conditioning in
the expressions for notational simplicity.}. Furthermore, the relays
that perform \cf\ have \csi\ about all the channels in the system.
This can happen at a negligible cost by proper feedback. We will
explain why we need this information when we discuss the \cf\
protocol in detail in Section~\ref{sec:MultiAntennaSingleRelay}. The
source(s) does not have instantaneous \csi. We also assume the
system is delay-limited and requires constant-rate transmission. We
note that under this assumption, information outage probability is
still well-defined and \dmt\ is a relevant performance
metric~\cite{CaireTB99}. There is also short-term average power
constraint that the transmitters have to satisfy for each codeword
transmitted. For more information about the effect of \csi\ at the
transmitter(s) and variable rate transmission on \dmt\ we refer the
reader to~\cite{KhoshnevisSabharwal04}.

\section{Preliminaries}\label{sec:preliminaries}
In this section we first introduce the notation, and present some
results we will use frequently in the paper.

For notational simplicity we write $f(\snr) \dot{=} \snr^c$, if
\[ \lim_{\snr \rightarrow \infty} \frac{\log f(\snr)}{\log \snr} =
c.\] The inequalities $\dot{\geq}$ and $\dot{\leq}$ are defined
similarly. In the rest of the paper $\mathbf{I}_i$ denotes the
identity matrix of size $i \times i$, $\dag$ denotes conjugate
transpose, and $|.|$ denotes the determinant operation. To clarify
the variables, we would like to note that $R_i$ denotes the $i$th
relay, whereas $R^{(.)}$ denotes transmission rates; e.g.
$R^{(T)}$ will be used for target data rate.

Let $R^{(T)}(\snr)$ denote the transmission rate of the system and
$P_e({\snr})$ denote the probability of error. Then we define
multiplexing gain $r$ and corresponding diversity $d(r)$ as
\begin{eqnarray*}
\lim_{\snr \rightarrow \infty}\frac{ R^{(T)}(\snr)}{\log \snr} &=
&r,
\\
\lim_{\snr \rightarrow \infty}\frac{ \log P_e(\snr)}{\log \snr} &=&
-d(r).
\end{eqnarray*}
The \dmt\ of an $m \times n$ \mimo\ is given by $d_{mn}(r)$, the
best achievable diversity, which is a piecewise-linear function
connecting the points $(k, d_{mn}(k))$, where $d_{mn}(k) =
(m-k)(n-k)$, $k=0,1,..., \min \{m,n\}$~\cite{ZhengT03}. Note that
$d_{mn}(r)=d_{nm}(r)$.

In~\cite{ZhengT03}, the authors prove that the probability of error
is dominated by the probability of outage. Therefore, in the rest of
the paper we will consider outage probabilities only.

We know that for any random channel matrix $\bH$ of size $n \times
m$ and for any input covariance matrix $Q$ of size $m \times m$
\cite{ZhengT03},
\begin{eqnarray}
\lefteqn{\sup_{Q \geq 0, \mathrm{trace}\{Q\}\leq {m \snr}} \log
\left| \mathbf{I}_n + \bH Q \bH^{\dag}\right|}  \nonumber \\ &\leq&
\log \left| \mathbf{I}_n+m\snr\bH
  \bH^{\dag}\right|. \label{eqn:covarianceremoval}
\end{eqnarray} Combined with the fact that a constant scaling in the transmit
power levels do not change the \dmt\ \cite{ZhengT03}, this bound
will be useful to establish \dmt\ results.

In a general multi-terminal network, node $k$ sends information to
node $l$ at rate \[ R^{(kl)} = \frac{1}{\eta}I(W_{kl};\hat{W}_{kl}),
\] where $\eta$ is the number of channel uses, $W_{kl}$ denotes the message for node $l$ at node $k$, and
$\hat{W}_{kl}$ is $W_{kl}$'s estimate at node $l$. Then the maximum
rate of information flow from a group of sources to a group of sinks
is limited by the minimum cut~\cite[Theorem 14.10.1]{Cover} and we
cite this result below.
\begin{proposition}\label{thm:CutsetBound}
Consider communication among $m$ nodes in a network. Let
$\mathcal{C}_i \subset \{1,2,...,m \}$ and $\mathcal{C}_i^c$ be the
complement of $\mathcal{C}_i$ in the set $\{ 1,2,...,m\}$. Also
$\bX^{(\mathcal{C}_i)}$ and $\bX^{(\mathcal{C}_i^c)}$ denote
transmitted signals from the sets $\mathcal{C}_i$ and
$\mathcal{C}_i^c$ respectively. $\bY^{(\mathcal{C}_i^c)}$ denotes
the signals received in the set $\mathcal{C}_i^c$. For information
rates $R^{(kl)}$ from node $k$ to $l$, there exists some joint
probability distribution $p(x_1,x_2,...,x_m)$, such that
\[ \sum_{k \in \mathcal{C}_i, l \in \mathcal{C}_i^c }R^{(kl)} \leq
I_{\mathcal{C}_i} =
I(\bX^{(\mathcal{C}_i)};\bY^{(\mathcal{C}_i^c)}|\bX^{(\mathcal{C}_i^c)}),\]
for all $\mathcal{C}_i \subset \{1,2,...,m \}$. Thus the total rate
of flow of information across cut-sets is bounded by the conditional
mutual information across that cut-set.
\end{proposition}
\vspace{0.2 cm} We can use the above proposition to find \dmt\ upper
bounds. Suppose $R^{(T_{kl})} = r^{(kl)} \log \snr$ denotes the
target data rate from node $k$ to node $l$, and $r^{(kl)}$ is its
multiplexing gain, $R^{(T_{\mathcal{C}_i})}=\sum_{k \in
\mathcal{C}_i, l \in \mathcal{C}_i^c }R^{(T_{kl})}$ denotes the sum
target data rate across cut-set $\mathcal{C}_i$ and
$r^{(\mathcal{C}_i)} = \sum_{k \in \mathcal{C}_i, l \in
\mathcal{C}_i^c }r^{(kl)}$ is its sum multiplexing gain. We say the
link from $k$ to $l$ is in outage if the event \[ \mathcal{E}_{kl} =
\{ R^{(kl)} < R^{(T_{kl})} \}\] occurs. Furthermore, the network
outage event is defined as
\[ \mathcal{E}_N = \bigcup_{k,l  \in  \{ 1,2,...,m\}, k \neq l}
\mathcal{E}_{kl},
\] which means the network is in outage if any link is in outage. Minimum network
outage probability is the minimum value of $P(\mathcal{E}_N)$ over
all coding schemes for the network. We name the \snr\ exponent of
the minimum network outage probability as \textit{maximum network
diversity}, $d(\bar{r})$, where $\bar{r}$ is a vector of all
$r^{(kl)}$'s. Then we have the following lemma, which says that the
maximum network diversity is upper bounded by the minimum diversity
over any cut.

\begin{lemma}\label{Lemma:CutsetUpperBound} For each $\mathcal{C}_i \subset
\{1,2,...,m \}$, define the maximum diversity order for that cut-set
$d_{\mathcal{C}_i}(r^{(\mathcal{C}_i)})$ as
\[ d_{\mathcal{C}_i}(r^{(\mathcal{C}_i)}) = - \lim_{\snr \rightarrow \infty} \frac{\log \min_{p(x_1,x_2,...,x_m)}P(I_{\mathcal{C}_i} <
R^{(T_{\mathcal{C}_i})})}{\log \snr}.\] Then the maximum network
diversity $d(\bar{r})$ is upper bounded as
\[ d(\bar{r}) \leq \min_{i } \{d_{\mathcal{C}_i}(r^{(\mathcal{C}_i)})\}.\]
\end{lemma}
\begin{proof} We provide the proof in Appendix~\ref{app:LemmaCutsetBound}. \end{proof}

In addition to Lemma~\ref{Lemma:CutsetUpperBound}, the following two
results will also be useful for some of the proofs.

\begin{lemma}[{\cite{HornJ85}}]\label{thm:Minkowski}
For two $n \times n$ positive definite matrices $A$ and $B$, if
$A-B$ is positive semi-definite, then $|A| \geq |B|$.
\end{lemma}
\vspace{0.2 cm}
\begin{lemma}\label{Lemma:AuxIneq}For two real numbers $x,y>0$, $xy/(x+y)<c$, where $c$ is a
non-negative real number, implies $x<2c$, or $y<2c$. Therefore, for
two non-negative random variables $X$ and $Y$,  $P\left(
{XY}/(X+Y)<c\right) \leq P(X<2c)+P(Y<2c)$.
\end{lemma}\vspace{0.2 cm}
The proof follows from simple arithmetic operations, which we omit
here.

\section{Multiple Antenna Nodes, Single Full-Duplex Relay}\label{sec:MultiAntennaSingleRelay}
\begin{figure}[t]
\centering
\includegraphics[width=2.2in]{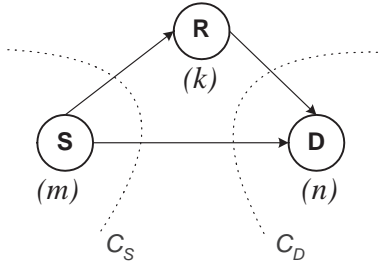}
\caption{System model for \textit{Problem 1}, the single relay
system. The source, the destination and the relay have $m$, $n$ and
$k$ antennas respectively.} \label{fig:MultiAntennaSingleRelay}
\end{figure}

The general multiple antenna, multiple source, destination, relay
network includes the multiple antenna relay channel consisting of a
single source, destination and relay, as a special case. Any attempt
to understand the most general network requires us to investigate
the multiple antenna relay channel in more detail. Therefore, in
this section we study \textit{Problem 1}, in which the source, the
destination and the relay has $m$, $n$ and $k$ antennas
respectively. This is shown in
Fig.~\ref{fig:MultiAntennaSingleRelay}. As clustering has a
significant effect on the \dmt\ performance of the network, we will
look into the non-clustered and clustered cases and examine the \df\
and \cf\ protocols. In this section the relay is full-duplex,
whereas in Section~\ref{sec:MultiAntennaSingleRelayHD}, the relay
will be half-duplex.

\subsection{Non-Clustered}\label{subsec:MultiAntennaSingleRelayNonClus}
Denoting the source and relay transmitted signals as $\bX_S$ and
$\bX_R$, when the system is non-clustered, the received signals at
the relay and at the destination are
\begin{eqnarray}
\bY_R &=& \bH_{SR}\bX_S + \bZ_R \label{eqn:MultiAntennaSingleRelayNonClusYR}\\
\bY_D &=& \bH_{SD}\bX_S + \bH_{RD}\bX_R +
\bZ_D,\label{eqn:MultiAntennaSingleRelayNonClusYD}
\end{eqnarray}
where $\bZ_R$ and $\bZ_D$ are the independent complex Gaussian
noise vectors at the corresponding node. $\bH_{SR}$, $\bH_{SD}$
and $\bH_{RD}$ are the $k \times m$, $ n \times m$ and $n \times
k$ channel gain matrices between the source and the relay, the
source and the destination, and the relay and the destination
respectively.
\begin{theorem}\label{thm:MultiAntennaSingleRelayNonClus}
The optimal \dmt\ for the non-clustered system of
Fig.~\ref{fig:MultiAntennaSingleRelay}, $d(r)$, is equal to
\begin{eqnarray*}
d(r) = \min \{d_{m(n+k)}(r),d_{(m+k)n}(r)\},
\label{eqn:MultiAntennaSingleRelayNonClusDMT}
\end{eqnarray*}
and the \cf\ protocol achieves this optimal \dmt\ for any $m$, $n$
and $k$.
\end{theorem}
\vspace{0.2 cm}
\begin{proof}

{\noindent{\textit{1) Upper Bound:}}} The instantaneous cut-set
mutual information expressions for cut-sets $\mathcal{C}_S$ and
$\mathcal{C}_D$ are
\begin{eqnarray}
I_{\mathcal{C}_S} &= &  I(\bX_S;\bY_R\bY_D|\bX_R) \label{eqn:MultiAntennaSingleRelayCutsetS} \\
I_{\mathcal{C}_D} &= & I(\bX_S\bX_R;\bY_D).
\label{eqn:MultiAntennaSingleRelayCutsetD}
\end{eqnarray}
To maximize these mutual information expressions we need to choose
$\bX_S$ and $\bX_R$ complex Gaussian with zero mean and covariance
matrices having trace constraints $P_S$ and $P_{R}$ respectively,
where $P_S$ and $P_{R}$ denote the average power constraints each
node has~\cite{TseV05}. Moreover, the covariance matrix of $\bX_S$
and $\bX_{R}$ should be chosen appropriately to maximize
$I_{\mathcal{C}_{D}}$. Then using (\ref{eqn:covarianceremoval}) to
upper bound $I_{\mathcal{C}_{S}}$ with $I_{\mathcal{C}_{S}}^\prime$
and $I_{\mathcal{C}_{D}}$ with $I_{\mathcal{C}_{D}}^\prime$ we can
write
\begin{eqnarray}
I_{\mathcal{C}_S} & \leq & I_{\mathcal{C}_S}' = \log K_{S,RD}^{\prime} \label{eqn:MultiAntennaSingleRelayCutset1}\\
I_{\mathcal{C}_D} & \leq & I_{\mathcal{C}_D}'  = \log K_{SR,D},
\label{eqn:MultiAntennaSingleRelayCutset2}
\end{eqnarray}where
\begin{eqnarray}
K_{S,RD}^{\prime} &\triangleq & \left|\mathbf{I}_{k+n} +
  \bH_{S,RD}\bH_{S,RD}^{\dag}P_S\right| \label{eqn:K_S,RD'}\\
K_{SR,D} & \triangleq & \left|\mathbf{I}_n+
\bH_{SR,D}\bH_{SR,D}^{\dag}(P_S+P_R)
 \right|, \label{eqn:K_SR,D}
\end{eqnarray}
with \begin{eqnarray} \bH_{S,RD} = \left[\begin{array}{c}
  \bH_{SR} \\
  \bH_{SD} \\
\end{array} \right],~~\bH_{SR,D} = \left[\begin{array}{cc}
  \bH_{SD} & \bH_{RD} \\
\end{array}\right].\label{eqn:H_S_RD}\end{eqnarray}

The above bounds suggest that the \csi\ at the relay does not
improve the \dmt\ performance under short term power constraint and
constant rate operation. The best strategy for the relay is to
employ beamforming among its antennas. For an $m$ antenna \mimo,
with total transmit power $P$, the beamforming gain can at most be
$mP$~\cite{ZhengT03}, which results in the same \dmt\ as using power
$P$. Therefore, \csi\ at the relay with no power allocation over
time does not improve the \dmt, it has the same the \dmt\ when only
receiver \csi\ is present.

Note that $P(I_{\mathcal{C}_i}^{\prime}) \dot{=}
\snr^{-d_{\mathcal{C}_i}^\prime(r)}$, $i= S,D$, with
$d_{\mathcal{C}_S}^\prime(r) = d_{m(n+k)}(r)$ and
$d_{\mathcal{C}_D}^\prime(r) = d_{(m+k)n}(r)$. Then using
Lemma~\ref{Lemma:CutsetUpperBound}, one can easily upper bound the
system \dmt\ by
\[ d(r) \leq \min \{d_{m(n+k)}(r),d_{(m+k)n}(r)\},\] for a target data rate $R^{(T)} = r \log
\snr$.

{\noindent{\textit{2) Achievability:}}} To prove the \dmt\ upper
bound of Theorem~\ref{thm:MultiAntennaSingleRelayNonClus} is
achievable, we assume the relay does full-duplex \cf\ as we explain
below. We assume the source, and the relay perform block Markov
superposition coding, and the destination does backward
decoding~\cite{KramerGG04,ZengKBuzo89,WillemsvM85}. The encoding is
carried over $B$ blocks, over which the fading remains fixed. In the
\cf\ protocol the relay performs Wyner-Ziv type compression with
side information taken as the destination's received signal. For
this operation the relay needs to know all the channel gains in the
system.

For the \cf\ protocol, as suggested in~\cite{CoverElG79} and
\cite{KramerGG04}, the relay's compression rate has to satisfy
\begin{eqnarray}
I(\hat{\bY}_R;\bY_R|\bX_R\bY_D)\leq I(\bX_R;\bY_D),
\label{eqn:CompConstr}
\end{eqnarray}
in order to forward $\hat{\bY}_R$ reliably to the destination. Here
$\hat{\bY}_R$ denotes the compressed signal at the relay. The
destination can recover the source message reliably if the
transmission rate $R^{(T)}=r \log \snr$ of the source is less than
the instantaneous mutual information $R^{(CF)}$
\begin{eqnarray}
R^{(CF)} = I(\bX_S;\hat{\bY}_R\bY_D|\bX_R).
\label{eqn:MultiAntennaSingleRelayCFrate}
\end{eqnarray}

We assume $\bX_S$ and $\bX_R$ are chosen independently, and have
covariance matrices $\mathbf{I}_m P_S/m$ and $\mathbf{I}_kP_R/k$
respectively. Also $\hat{\bY}_R=\bY_R+\hat{\bZ}_R$, where
$\hat{\bZ}_R$ is a length $k$ vector with complex Gaussian random
entries with zero mean. $\hat{\bZ}_R$ has covariance matrix
$\hat{N}_R\mathbf{I}_k$, and its entries are independent from all
other random variables. We define
\begin{eqnarray}
L_{S,D}  &\triangleq&  \left| \bH_{SD}\bH_{SD}^{\dagger}\frac{P_S}{m}+\mathbf{I}_n \right| \label{eqn:L_S,D} \\
L_{SR,D} &\triangleq &
\left|\bH_{SD}\bH_{SD}^{\dag}\frac{P_S}{m}+\bH_{RD}\bH_{RD}^{\dag}\frac{P_R}{k}+\mathbf{I}_n
\right|, \label{eqn:L_SR,D} \\
L_{S,RD} & \triangleq& \left|
\bH_{S,RD}\bH_{S,RD}^{\dag}\frac{P_S}{m}+\left[ \begin{array}{cc}
(\hat{N}_R+1)\mathbf{I}_k & \mathbf{0} \\ \mathbf{0} & \mathbf{I}_n
\end{array}\right]\right|.\label{eqn:L_S,RD}\\
L_{S,RD}^{\prime} &\triangleq &
\left|{\bH_{S,RD}}\bH_{S,RD}^{\dag}\frac{P_S}{m}+\mathbf{I}_{k+n}\right|
\label{eqn:L_S,RD'}
\end{eqnarray}
Then we have
\begin{eqnarray*}
I(\hat{\bY}_R;\bY_R|\bX_R\bY_D) &=& \log
\frac{L_{S,RD}}{L_{S,D}\hat{N}_R^k}\\
I(\bX_R;\bY_D) &=& \log \frac{L_{SR,D}}{L_{S,D}}.
\end{eqnarray*}
To satisfy the compression rate constraint in
(\ref{eqn:CompConstr}), using the \csi\ available to it, the relay
ensures that the compression noise variance $\hat{N}_R$ satisfies
$\hat{N}_R=\sqrt[k]{L_{S,RD}/L_{SR,D}}$. Note that both sides of
this equation are functions of $\hat{N}_R$. Then
\begin{eqnarray}
R^{(CF)} & = &I(\bX_S;\hat{\bY}_R\bY_D|\bX_R) \nonumber \\
 &=& \log \frac{L_{S,RD}}{\left(\sqrt[k]{\frac{L_{S,RD}}{L_{SR,D}}}+1\right)^k}  \nonumber\\
&=& \log \left(
\frac{\sqrt[k]{L_{S,RD}}\sqrt[k]{L_{SR,D}}}{\sqrt[k]{L_{S,RD}}+\sqrt[k]{L_{SR,D}}}
\right)^k. \label{eqn:MultiAntennaSingleRelayCFachievable}
\end{eqnarray}

To prove the \dmt\ of
(\ref{eqn:MultiAntennaSingleRelayCFachievable}) we need to find how
probability of error decays with increasing \snr\ when the target
rate increases as $R^{(T)} = r\log \snr$. As the error events are
dominated by outage events, we use the following bound on the
probability of outage
\begin{eqnarray}
\lefteqn{P(\mbox{outage at D})} \nonumber\\
 &=& P\left(I(\bX_S;\hat{\bY}_R\bY_D|\bX_R)<r\log \snr\right) \label{eqn:MultiAntennaSingleRelayNonClusPout1}\\
&=& P\left(
\frac{\sqrt[k]{L_{S,RD}}\sqrt[k]{L_{SR,D}}}{\sqrt[k]{L_{S,RD}}+\sqrt[k]{L_{SR,D}}}
<
\snr^{\frac{r}{k}} \right) \\
&\overset{(a)}{\leq}& P \left(
\frac{\sqrt[k]{L_{S,RD}'}\sqrt[k]{L_{SR,D}}}{\sqrt[k]{L_{S,RD}'}+\sqrt[k]{L_{SR,D}}}
<
 \snr^{\frac{r}{k}} \right) \\
 &\overset{(b)}{\leq}& P \left( \sqrt[k]{L_{S,RD}'} < 2\snr^{\frac{r}{k}} \right) \nonumber \\
 &&{+}\: P \left( \sqrt[k]{L_{SR,D}} < 2\snr^{\frac{r}{k}} \right) \\
& = & P \left( L_{S,RD}' < 2^k\snr^{r} \right) \nonumber
\\
&&{+}\: P \left( L_{SR,D} < 2^k\snr^{r} \right) \label{eqn:MultiAntennaSingleRelayNonClusPout2}\\
& \overset{(c)}{\dot{=}} & {\snr^{-d_{\mathcal{C}_S}'(r)}}+{\snr^{-d_{\mathcal{C}_D}'(r)}} \label{eqn:MultiAntennaSingleRelayNonClusPout3} \\
& {=} &{\snr^{-d_{m(n+k)}(r)}} +{\snr^{-d_{(m+k)n}(r)}}. \nonumber
\end{eqnarray}
where for $(a)$ we first used Lemma~\ref{thm:Minkowski} to show
$L_{S,RD} \geq L_{S,RD}^{\prime} $ and the fact that the ratio
$xy/(x+y)$ is monotone decreasing with decreasing $x$ for $x, y>0$,
$(b)$ follows from Lemma~\ref{Lemma:AuxIneq}, and $(c)$ follows
because $\log L_{S,RD}'$ and $\log L_{SR,D}$ are same as the cut-set
mutual information expressions $I_{\mathcal{C}_{S}}$ and
$I_{\mathcal{C}_D}$ except a constant scaling factor of \snr, and a
constant scaling in \snr\ in the probability expression does not
change the diversity gain. We conclude that the system \dmt\
$d_{CF}(r) \geq \min \{d_{m(n+k)}(r),d_{(m+k)n}(r) \}$. This result
when combined with the upper bound results in
\begin{eqnarray*}
d_{CF}(r) = \min \{d_{m(n+k)}(r),d_{(m+k)n}(r).
\}\label{eqn:MultiAntennaNonClusCFdmt}
\end{eqnarray*}
\end{proof}

As an alternative to the \cf\ protocol, the relay can use the \df\
protocol. When the source, the destination and the relay all have a
single antenna each, it is easy to show that the \df\ protocol also
achieves the \dmt\ upper bound, which is equal to $d_{12}(r)$. The
following theorem derives the \dmt\ of the \df\ protocol for
arbitrary $m$, $n$ and $k$ and shows that the optimality of \df\
does not necessarily hold for all $m$, $n$ and $k$.

\begin{theorem}\label{thm:MultiAntennaSingleRelayNonClusDF} For the system in
Fig.~\ref{fig:MultiAntennaSingleRelay}, \df\ achieves the \dmt\
\begin{eqnarray}
d_{DF}(r) = \left\{\begin{array}{l}
  \min \{d_{(m+k)n}(r),d_{mn}(r)+d_{mk}(r)\} \\  \mbox{~~~~~~~~~~~~~~~~~if~}  0\leq r\leq \min \{m,n,k\}\\
  d_{mn}(r)  \\ \mbox{~~~~~if~}  \min\{m,n,k\} < r\leq \min \{m,n\}\\
\end{array} \right..\label{eqn:MultiAntennaNonClusDFdmt}
\end{eqnarray}
\end{theorem}
\vspace{0.2 cm}
\begin{proof}
We provide the proof in
Appendix~\ref{app:MultiAntennaSingleRelayNonClusDF}.
\end{proof}

We next consider examples for the \df\ \dmt\ performance and compare
with Theorem~\ref{thm:MultiAntennaSingleRelayNonClus}. If $m$ or $n$
(or both) is equal to 1, we find that \df\ meets the bound in
Theorem~\ref{thm:MultiAntennaSingleRelayNonClus} and is optimal
irrespective of the value of $k$. Similarly we can show that for
cases such as $(m,n,k)=(3,2,2)$ or $(m,n,k)=(4,2,3)$, as
$d_{(m+k)n}(r)< d_{mn}(r)+d_{mk}(r)$ for all $r$, \df\ is optimal. A
general necessary condition for \df\ to be optimal for all
multiplexing gains is $m \geq n$. If $m<n$, then
$d_{mn}(r)+d_{mk}(r) \leq d_{m(n+k)}(r) < d_{(m+k)n}$, and \df\ will
be suboptimal.

Whenever $\min\{m,n,k\} = k$, the degrees of freedom in the direct
link is larger than the degrees of freedom in the source to relay
link, that is $\min\{m,n\} \geq \min\{ m,k\}$. For multiplexing
gains in the range $\min\{m,n,k\} < r \leq \min \{m,n\}$, the relay
can never help and the system has the direct link \dmt\ $d_{mn}(r)$.
Therefore, \df\ loses its optimality. For example, if $(m,n,k) =
(3,2,1)$, then \df\ is optimal only for multiplexing gains up to
$1/2$, but for $1/2 \leq r \leq 2$, \df\ is suboptimal. In
particular, \df\ does not improve upon $d_{32}(r)$ in the range $1
\leq r \leq 2$.

\begin{figure}[t]
\centering
\includegraphics[width=3.4in]{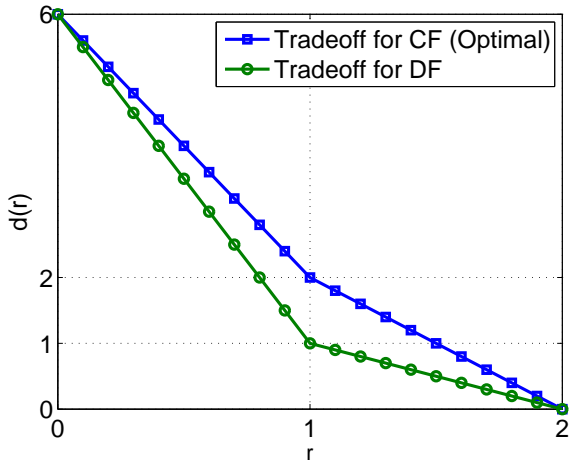}
\caption{The source has 2, the destination has 2, and the relay has
1 antenna, $(m,n,k)=(2,2,1)$. The network is non-clustered.}
\label{fig:MultiAntennaSingleRelayDMT1_NonClus}
\end{figure}

\begin{figure}[t]
\centering
\includegraphics[width=3.4in]{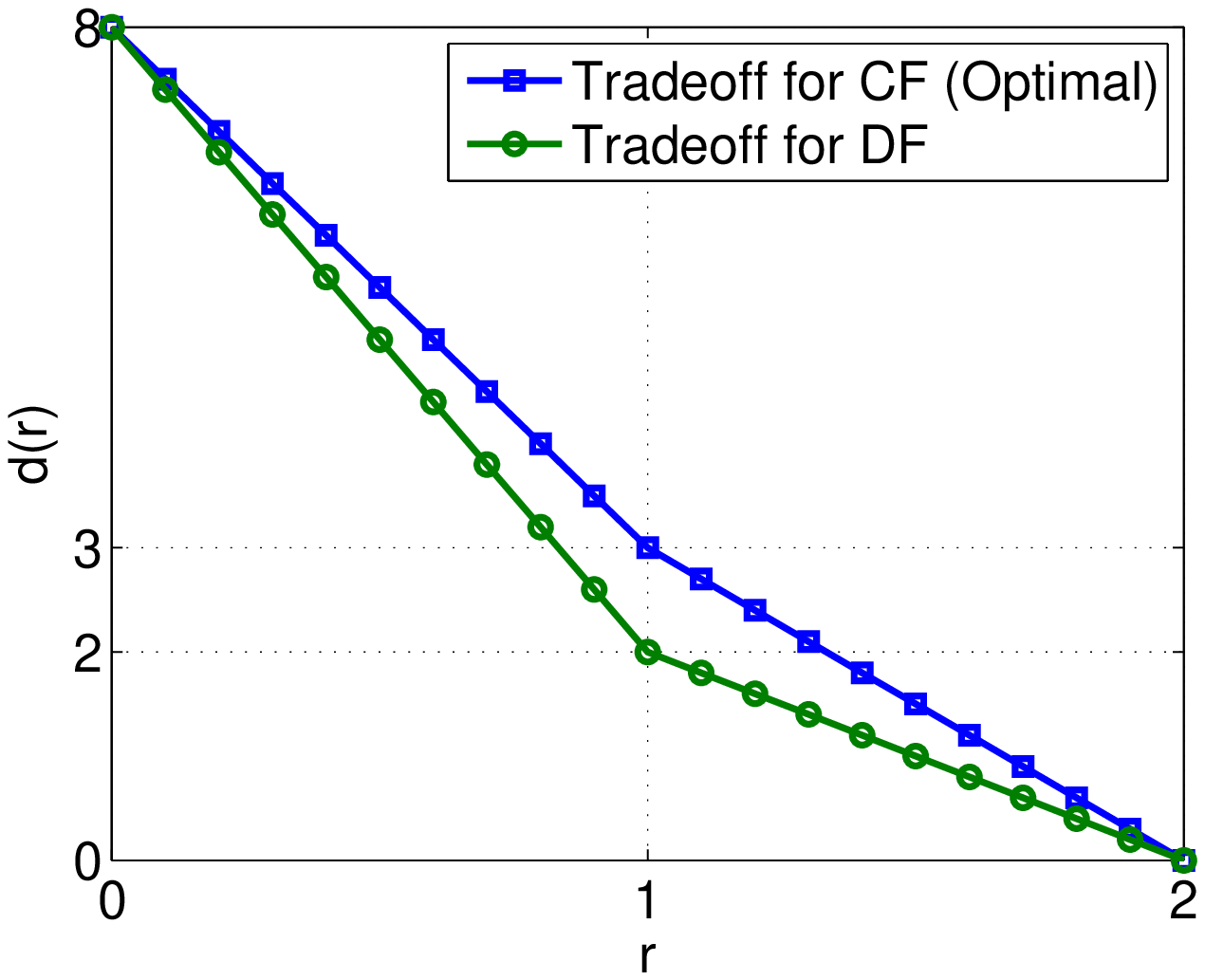}
\caption{The source has 2, the destination has 2, and the relay has
2 antennas, $(m,n,k)=(2,2,2)$. The network is non-clustered.}
\label{fig:MultiAntennaSingleRelayDMT2_NonClus}
\end{figure}

\begin{figure}
\centering
\includegraphics[width=3.4in]{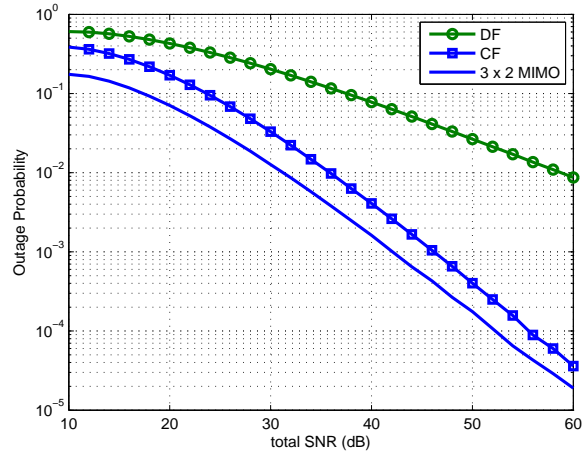}
\caption{Outage probability versus total \snr\ for the non-clustered
multiple-antenna, single full-duplex relay system,
$(m,n,k)=(2,2,1)$, $r = 1.5$.} \label{fig:SRD_212_NonClustered_FD}
\end{figure}

Fig.~\ref{fig:MultiAntennaSingleRelayDMT1_NonClus} shows the \cf\
and \df\ \dmt\ for $(m,n,k)=(2,2,1)$, and
Fig.~\ref{fig:MultiAntennaSingleRelayDMT2_NonClus} shows the \cf\
and \df\ \dmt\ for $(m,n,k)=(2,2,2)$. When we compare the figures,
we see that the \cf\ protocol is always \dmt\ optimal, but the \df\
protocol can still be suboptimal even when the source to relay link
has the same degrees of freedom as the link from the source to the
destination. The suboptimal behavior of \df\ arises because the
outage event when the relay cannot decode can dominate for general
$m,n$ and $k$. In addition to this, for multiplexing gains larger
than $\min \{ m,n,k\}$, the relay never participates in the
communication because it is degrees of freedom limited and cannot
decode large multiplexing gain signals. For this region, we observe
the direct link behavior. We conclude that soft information
transmission, as in the \cf\ protocol, is necessary at the relay not
to lose diversity or multiplexing gains.

Fig.~\ref{fig:SRD_212_NonClustered_FD} shows the outage probability
versus total \snr\ for \df\ and \cf\ protocols for
$(m,n,k)=(2,2,1)$, $R^{(T)} = r \log (P_S+P_R)$, $r = 1.5$. The
channel gain matrices $\bH_{SR}$, $\bH_{SD}$ and $\bH_{RD}$ have
i.i.d. complex Gaussian entries, with real and imaginary parts zero
mean and variance $1/2$ each. We have $P_S = 2 P_R$. The figure also
includes the $3 \times 2$ \mimo\ for comparison. We assume the total
power constraint is the same for both the \mimo\ and relay systems.
In the \mimo\ system the antennas share the total power equally and
send uncorrelated signals. We observe that while \df\ achieves
$d=0.5$, \cf\ achieves $d=1$ and performs similar to $3 \times 2$
\mimo\ as predicted by
Theorem~\ref{thm:MultiAntennaSingleRelayNonClus}.

The above analysis also reveals that \cf\ and \df\ protocols do not
always behave similar, unlike the single antenna relay system. The
degrees of freedom available also has an effect on relaying
strategies.

\subsection{Clustered}\label{subsec:MultiAntennaSingleRelayClus}
Clustering can sometimes improve the system performance, since it
eliminates fading between some of the users. We will observe an
example of this in Section~\ref{sec:SingleAntenna}, when there are
multiple relays. Therefore, in this subsection we study the \dmt\
behavior of a single relay system, when the relay is clustered with
the source. The analysis presented in this subsection can easily be
modified if relay is clustered with the destination. The system
input and output signals are same as
(\ref{eqn:MultiAntennaSingleRelayNonClusYR}) and
(\ref{eqn:MultiAntennaSingleRelayNonClusYD}) but for the clustered
case all the entries of $\bH_{SR}$ are equal to $\sqrt{G}$.

\begin{theorem}\label{thm:MultiAntennaSingleRelayClusCF}
For the system in Fig.~\ref{fig:MultiAntennaSingleRelay} when the
relay is clustered with the source, the \cf\ protocol is optimal
from the \dmt\ perspective for all $(m,n,k)$.
\end{theorem}\vspace{0.2 cm}

We omit the proof, as the achievability follows the same lines as in
Theorem~\ref{thm:MultiAntennaSingleRelayNonClus}, and results in the
same outage probability expression in
(\ref{eqn:MultiAntennaSingleRelayNonClusPout3}), which is equal to
the upper bound.

We next compute the \dmt\ of the clustered system explicitly for
$m=1,2$, arbitrary $n$ and $k$. We conjecture the same form holds
for arbitrary $m$ as well.
\begin{theorem}\label{thm:MultiAntennaSingleRelayClusUB}
For the clustered system of Fig.~\ref{fig:MultiAntennaSingleRelay},
for $(m,n,k)$, $m=1,2$, the \dmt\ is given by
\[ d(r) = \left\{\begin{array}{l}
  d_{(m+k)n}(r)  \mbox{~~~~~~~~~~~~~~~~~~~~~~~~~~~if~} 0 \leq r < 1 \\
  \min \{d_{m(n+1)}(r), d_{(m+k)n}(r)\}  \\ \mbox{~~~~~~~~~~~~~~~~~~~~~~~~~~~~~~if~} 1 \leq r \leq \min\{ m,n\} \\
\end{array} \right. .\]
\end{theorem}
\vspace{0.2 cm}
\begin{proof}
We provide the proof in
Appendix~\ref{app:MultiAntennaSingleRelayClusteredUB}.
\end{proof}
For the clustered case, for any $m$ or $k$, $\bH_{SR}$ has rank 1,
hence we have the following conjecture.
\begin{conjecture} Theorem~\ref{thm:MultiAntennaSingleRelayClusUB}
is true for arbitrary $(m,n,k)$.
\end{conjecture}
\vspace{0.2 cm}

We observe that if $m=2$, although the source and the relay both
have multiple antennas, as the channel gain matrix in between is
\awgn\ and has rank 1, it can only support multiplexing gains up to
1. This is because having multiple antennas at the transmitter
and/or the receiver in an \awgn\ channel only introduces power gain.
Therefore, $I_{\mathcal{C}_S}$, the mutual information across
cut-set $\mathcal{C}_S$, never results in outage for multiplexing
gains up to 1. For multiplexing gains $r \geq 1$, this cut-set
results in a \dmt\ of $d_{2(n+1)}(r)$, even though the relay has $k$
antennas. The next theorem is a counterpart of
Theorem~\ref{thm:MultiAntennaSingleRelayNonClusDF} for the clustered
case.

\begin{figure}[t]
\centering
\includegraphics[width=3.4in]{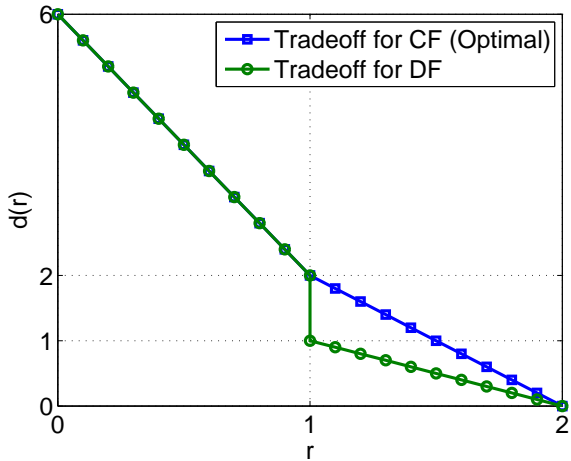}
\caption{The source has 2, the destination has 2, and the relay has
1 antenna, $(m,n,k)=(2,2,1)$. The relay is clustered with the
source.} \label{fig:MultiAntennaSingleRelayDMT1_Clus}
\end{figure}

\begin{figure}[h]
\centering
\includegraphics[width=3.4in]{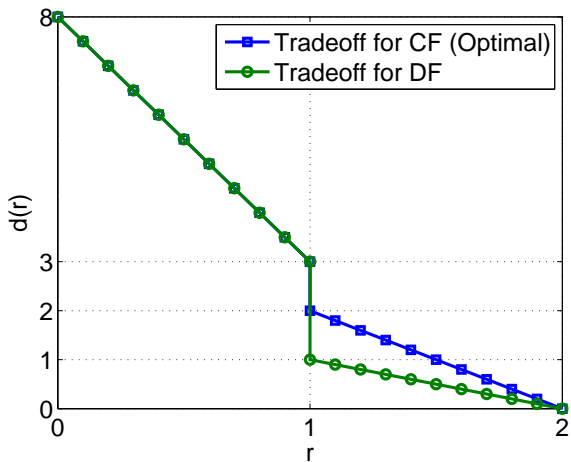}
\caption{The source has 2, the destination has 2, and the relay has
2 antennas, $(m,n,k)=(2,2,2)$. The relay is clustered with the
source.} \label{fig:MultiAntennaSingleRelayDMT2_Clus}
\end{figure}

\begin{theorem} For the system in Fig.~\ref{fig:MultiAntennaSingleRelay},
when the relay is clustered with the source, the \df\ protocol achieves the \dmt\
\begin{eqnarray*} d_{DF}(r) &=&\left\{ \begin{array}{lll}
  d_{(m+k)n}(r) & \mbox{if} & 0\leq r < 1\\
  d_{mn}(r) & \mbox{if} & 1 < r\leq \min \{m,n\}\\
\end{array} \right. .
\end{eqnarray*}
\end{theorem}
\vspace{0.2 cm}
\begin{proof}
The outage probability for \df\ is the same as
(\ref{eqn:MultiAntennaSingleRelayDFout}), in the non-clustered case
of Appendix~\ref{app:MultiAntennaSingleRelayNonClusDF}. If $0 \leq r
< 1$, then the probability that the relay is in outage is 0. On the
other hand, if $1 \leq r \leq \min \{ m,n\}$, the probability that
the relay can decode is 0, since the source-relay channel can only
support multiplexing gains up to 1.
\end{proof}

We have seen in Section~\ref{subsec:MultiAntennaSingleRelayNonClus}
that \df\ is in general suboptimal for non-clustered multi-antenna
relay channel. However, once we cluster the relay with the source,
there are no more outages in the source-relay channel for
multiplexing gains up to 1, and the \df\ performance improves in
this range. However, even with clustering \df\ does not necessarily
meet the \dmt\ upper bound for arbitrary $(m,n,k)$.

Fig.~\ref{fig:MultiAntennaSingleRelayDMT1_Clus} compares the
clustered \cf\ and \df\ \dmt\ for $(m,n,k)=(2,2,1)$, and
Fig.~\ref{fig:MultiAntennaSingleRelayDMT2_Clus} for
$(m,n,k)=(2,2,2)$. Comparing with the upper bound, we can see that
clustering improves the \df\ performance in the range $0 \leq r \leq
1$, where \df\ achieves the upper bound. However, for multiplexing
gains larger than 1, \df\ is still suboptimal. In fact, in this
range the relay can never decode the source even though they are
clustered and hence cannot improve the direct link performance.
Although clustering improves the \df\ performance for low
multiplexing gains, it is not beneficial for multiple antenna
scenarios in terms of \dmt, it can in fact decrease the optimal
diversity gain. This is because when two nodes have multiple
antennas, clustering decreases the degrees of freedom in between.
This can also be observed comparing
Theorem~\ref{thm:MultiAntennaSingleRelayClusUB} and
Theorem~\ref{thm:MultiAntennaSingleRelayNonClus}, as well as the
optimal strategies in
Fig.~\ref{fig:MultiAntennaSingleRelayDMT2_Clus} with
Fig.~\ref{fig:MultiAntennaSingleRelayDMT2_NonClus}. We will also
study the effects of clustering in single antenna multiple relay
scenarios in Section~\ref{sec:SingleAntenna}.

\section{Multiple Antenna Nodes, Single Half-Duplex Relay}\label{sec:MultiAntennaSingleRelayHD}
In the previous section, we studied the relay channel when the relay
is full-duplex. Although this is an ideal assumption about the
relay's physical capabilities, it helps us understand the
fundamental differences between the \df\ and \cf\ protocols. In this
section we assume a half-duplex, non-clustered relay to study how
this affects the \dmt\ behavior of the relay channel.

In half-duplex operation a state variable $Q$, which takes the value
$q_1$ if the relay is listening, or $q_2$ if the relay is
transmitting, controls the relay operation. For a more general
treatment that considers three different states depending on whether
the relay is in sleep, listen or talk states see~\cite{Kramer04}.
Our results in this section would also be applicable for this case
as well.

Depending on how the state $Q$ is designed, half-duplex protocols
can be \textit{random} or \textit{fixed}. In fixed protocols, the
state does not convey additional information to the destination via
the state random variable $Q$, whereas in random protocols the relay
breaks its transmission and reception intervals into small blocks to
send extra information through the state. This is equivalent to
considering the random binary state as a channel input and designing
code books to convey information through $Q$.

\begin{figure}[t]
\centering
\includegraphics[width=1.8in]{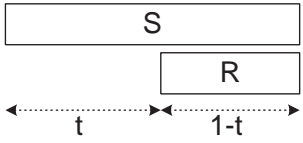} \caption{A fixed and static
half-duplex relaying protocol, where the relay listens for $t$
fraction of time, transmits for $(1-t)$ fraction, where $t$ is a
constant.} \label{fig:MultiAntennaSingleRelayHD}
\end{figure}

Another categorization based on the state variable $Q$ is
\textit{dynamic} versus \textit{static}. If the state is controlled
based on channel realizations, we have a dynamic protocol. On the
other hand, if $Q$ does not depend on \csi, the protocol is called
static. Note that fixed protocols are included in random ones, and
static protocols in dynamic ones. The most commonly used relaying
protocols are fixed and static, and of the form shown in
Fig.~\ref{fig:MultiAntennaSingleRelayHD}. The \ddf\ protocol
of~\cite{AzarianEGS05} is an example to a fixed, dynamic protocol.

For the multiple antenna half-duplex relay channel, using
Lemma~\ref{Lemma:CutsetUpperBound} directly, results in the
full-duplex bound, which is not tight for half-duplex operation.
Therefore, we first state the following lemma to provide a
half-duplex \dmt\ upper bound for random, static protocols. The
lemma also suggests that sending information through the state does
not improve \dmt. Lemma~\ref{Lemma:DMTboundHD} can be modified for
random, dynamic state protocols as well.
\begin{lemma}\label{Lemma:DMTboundHD} For the multiple antenna half-duplex relay channel,
the half-duplex \dmt\ upper bound for random, static state protocols
is equal to
\begin{eqnarray}
d(r) \leq \max_{p(q)} \min \{
d_{\mathcal{C}_S}(r,p(q)),d_{\mathcal{C}_D}(r,p(q)) \},
\end{eqnarray} where
\begin{eqnarray} \lefteqn{d_{\mathcal{C}_i}(r,p(q))} \nonumber \\
& =& -\lim_{\snr \rightarrow \infty} \frac{\log \min_{p(x_S,x_R|q)}
P(I_{\mathcal{C}_i}< R^{(T)})}{\log \snr}, \label{eqn:div_HD}
\end{eqnarray} $i = S,D$.
\end{lemma}
\vspace{0.2 cm}
\begin{proof} We provide the proof in
Appendix~\ref{app:LemmaDMTboundHD}.
\end{proof}

Our next theorem and corollary provide the first half-duplex \dmt\
achieving relaying protocol in the literature.

\begin{theorem}\label{thm:MultiAntennaSingleRelayHD_CF}
For the random, dynamic state, half-duplex relay channel with $m$
antenna source, $k$ antenna relay and $n$ antenna destination, the
\cf\ protocol is \dmt\ optimal.
\end{theorem} \vspace{0.2 cm}
\begin{corollary}\label{cor:HD_CF_111}
For $(m,n,k) =  (1,1,1)$, the half-duplex \dmt\ upper bound is equal
to the full-duplex \dmt, $2(1-r)$. Therefore, \cf\ is a \dmt\
optimal half-duplex protocol for the single antenna relay channel.
\end{corollary}
\vspace{0.2 cm}

\begin{proof}\textit{[Theorem~\ref{thm:MultiAntennaSingleRelayHD_CF}]} First, we prove that \cf\ is optimal among static protocols and then show
that the same proof follows for dynamic protocols as well.

At state $q_1$, the received signals at the relay and the
destination are
\begin{eqnarray*}
\bY_{R,1} & = & \bH_{SR} \bX_{S,1} + \bZ_{R,1} \\
\bY_{D,1} & = & \bH_{SD} \bX_{S,1} + \bZ_{D,1}
\end{eqnarray*} and at state $q_2$, the received signal at the destination is given
as
\begin{eqnarray*}
\bY_{D,2} & = & \bH_{SD} \bX_{S,2} + \bH_{RD} \bX_{R,2}+ \bZ_{D,2}.
\end{eqnarray*}
Here $\bX_{S,l}$, and $\bX_{R,l}$ are of size $m$, and $k$ column
vectors respectively and denote transmitted signal vectors at node
$S$, and $R$ at state $q_l$, $l= 1,2$. Similarly $\bY_{R,l}$ and
$\bY_{D,l}$ are the received signal vectors of size $k$ and $n$.

We first find an upper bound to the \dmt\ using
Lemma~\ref{Lemma:DMTboundHD}. Without loss of generality we use a
fixed state static protocol as shown in
Fig.~\ref{fig:MultiAntennaSingleRelayHD}. This is justified by the
proof of Lemma~\ref{Lemma:DMTboundHD}, which states that fixed and
random protocols have the same \dmt\ upper bound. For the
half-duplex relay channel using the cut-set $\mathcal{C}_S$ around
the source and $\mathcal{C}_D$ around the destination as shown in
Fig.~\ref{fig:MultiAntennaSingleRelay}, we have
\cite{KhojastepourSA03}
\begin{eqnarray}
I_{\mathcal{C}_S}(t) &=& t I(\bX_S;\bY_R \bY_D|q_1)\nonumber \\
&&{ +}\: (1-t)I(\bX_S;\bY_D|\bX_R,q_2)  \label{eqn:cutset1} \\
I_{\mathcal{C}_D}(t) &=& t I(\bX_S;\bY_D|q_1) \nonumber \\
&&{+}\: (1-t)I(\bX_S\bX_R;\bY_D|q_2). \label{eqn:cutset2}
\end{eqnarray}
We define
\begin{eqnarray}
K_{S,D}  &\triangleq&  \left| \bH_{SD}\bH_{SD}^{\dagger}P_S
+\mathbf{I}_n \right|. \label{eqn:K_S,D}
\end{eqnarray}
Then we can upper bound $I_{\mathcal{C}_S}(t)$ and
$I_{\mathcal{C}_D}(t)$ with $I_{\mathcal{C}_S}'(t)$ and
$I_{\mathcal{C}_D}'(t)$ as
\begin{eqnarray}
I_{\mathcal{C}_S}(t)\leq I_{\mathcal{C}_S}'(t) = t\log K_{S,RD}^{\prime}+(1-t)\log K_{S,D} \label{eqn:C_1(t)}\\
I_{\mathcal{C}_D}(t)\leq I_{\mathcal{C}_D}'(t) = t\log
K_{S,D}+(1-t)\log K_{SR,D}\label{eqn:C_2(t)}
\end{eqnarray}
where $K_{S,RD}^{\prime}$ and $K_{SR,D}$ are defined in
(\ref{eqn:K_S,RD'}) and (\ref{eqn:K_SR,D}).

For a target data rate $R^{(T)}=r\log\snr$, and for a fixed $t$, if
$P(I_{\mathcal{C}_i}'(t)<R^{(T)}) \dot{=}
\snr^{-d_{\mathcal{C}_i}'(r,t)}$, $i = S,D$, then
$d_{\mathcal{C}_i}(r,t)$ of Lemma~\ref{Lemma:DMTboundHD} satisfies
$d_{\mathcal{C}_i}(r,t) \leq d_{\mathcal{C}_i}'(r,t),$ where we
denoted $d_{\mathcal{C}_i}(r,p(q))$ with $d_{\mathcal{C}_i}(r,t)$
with an abuse of notation. Therefore, the best achievable diversity
for the half-duplex relay channel for fixed $t$ satisfies
\begin{equation}
d(r,t) \leq \min \{ d_{\mathcal{C}_S}'(r,t),
d_{\mathcal{C}_D}'(r,t)\}.\label{eqn:RC_DMT_upperbound}
\end{equation} Optimizing over $t$ we find an upper bound on the
static multiple antenna half-duplex relay channel \dmt\ as
\begin{equation}
d(r) \leq \max_{t}\min \{ d_{\mathcal{C}_S}'(r,t),
d_{\mathcal{C}_D}'(r,t)\}. \label{eqn:RC_DMT_upperbound_maximized}
\end{equation}

Appendix~\ref{app:MultiAntennaSingleRelayHD_CF} shows that
half-duplex \cf\ achieves the upper bound in
(\ref{eqn:RC_DMT_upperbound_maximized}). For dynamic protocols, the
\dmt\ upper bound will change because of the \csi\ available at the
relay. Appendix~\ref{app:MultiAntennaSingleRelayHD_CF} also shows
that if \cf\ is allowed dynamic operation, it achieves the dynamic
\dmt\ upper bound as well.
\end{proof}

\subsection{Static Half-Duplex \dmt\
Computation}\label{subsec:HD_DMT_Computation} In general it is hard
to compute the exact \dmt\ of
Theorem~\ref{thm:MultiAntennaSingleRelayHD_CF}. In particular for
static protocols, to find $d_{\mathcal{C}_S}'(r,t)$ and
$d_{\mathcal{C}_D}'(r,t)$ for general $m$, $n$ and $k$ we need to
calculate the joint eigenvalue distribution of two correlated
Hermitian matrices, $\bH_{SD}\bH_{SD}^\dag$ and
$\bH_{S,RD}\bH_{S,RD}^\dag$ or $\bH_{SD}\bH_{SD}^\dag$ and
$\bH_{SR,D}\bH_{SR,D}^\dag$. However, when $m=1$, both
$\bH_{SD}\bH_{SD}^\dag$ and $\bH_{S,RD}\bH_{S,RD}^\dag$ reduce to
vectors and it becomes easier to find $d_{\mathcal{C}_S}'(r,t)$.
Similarly, when $n=1$, $\bH_{SD}\bH_{SD}^\dag$ and
$\bH_{SR,D}\bH_{SR,D}^\dag$ are vectors, and
$d_{\mathcal{C}_D}'(r,t)$ can be found.

An explicit form for $d_{\mathcal{C}_S}'(r,t)$ is given in the
following theorem.
\begin{theorem}
For $m=1$, $d_{\mathcal{C}_S}'(r,t)$ is given as
\begin{eqnarray*}
d_{\mathcal{C}_S}'(r,t) = \left\{ \begin{array}{lll}
  n+k-k\frac{r}{t} & \mbox{if} & r\leq t,\mbox{~and~} t\leq \frac{k}{n+k} \vspace{0.15 cm}\\
  n\left(\frac{1-r}{1-t}\right) & \mbox{if} & r\geq t, \mbox{~and~} t \leq \frac{k}{n+k} \vspace{0.15 cm}\\
  (n+k)(1-r) & \mbox{if} & t \geq \frac{k}{n+k}
\end{array}\right..
\end{eqnarray*}
For $n=1$ and for arbitrary $m$ and $k$, $d_{\mathcal{C}_D}'(r,t)$
has the same expression as $d_{\mathcal{C}_S}'(r,t)$ if $n$ and
$t$ are replaced with $m$ and $(1-t)$ in the above expressions.
\label{thm:HalfDuplexDMTBound}
\end{theorem}
\vspace{0.2 cm}
\begin{proof} The proof follows the ideas presented
in~\cite{ZhengT03,AzarianEGS05}, and is provided in
Appendix~\ref{app:HalfDuplexDMTBound}.
\end{proof}

\begin{figure}[t]
\begin{centering}
\includegraphics[width=3.26 in]{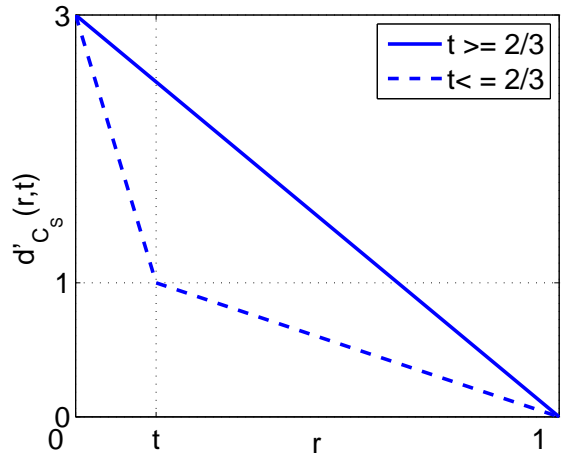}\\
\caption{\dmt\ upper bound for the cut-set around the source,
$\mathcal{C}_S$. The source has 1, the destination has 1, and the
relay has 2 antennas, $(m,n,k)=(1,1,2)$. The network is
non-clustered. Note that as $m=n$, $d_{\mathcal{C}_S}'(r,t) =
d_{\mathcal{C}_D}'(r,1-t)$. The upper bound in
(\ref{eqn:RC_DMT_upperbound_maximized}) reaches its maximum for
$t=1/2$. The solid line in the figure is also equal to the
full-duplex bound.}\label{fig:rc_m1k2n1a}
\end{centering}
\end{figure}

Although we do not have an explicit expression for
$d_{\mathcal{C}_S}'(r,t)$ or $d_{\mathcal{C}_D}'(r,t)$ for general
$(m,n,k)$, we can comment on some special cases and get insights
about multiple antenna, half-duplex behavior. First we observe that
$d_{\mathcal{C}_S}'(r,t)$ and $d_{\mathcal{C}_D}'(r,t)$ depend on
the choice of $t$, and the upper bound of
(\ref{eqn:RC_DMT_upperbound_maximized}) is not always equal to the
full-duplex bound. As an example consider $(m,n,k)=(1,1,2)$, for
which $d_{\mathcal{C}_S}'(r,t)$ is shown in
Fig.~\ref{fig:rc_m1k2n1a}. To achieve the full-duplex bound for all
$r$, $d_{\mathcal{C}_S}'(r,t)$ needs to have $t \geq 2/3$, whereas
$d_{\mathcal{C}_D}'(r,t)$ needs $t \leq 1/3$. As both cannot be
satisfied simultaneously, $d(r,t)$ will be less than the full-duplex
bound for all $t$.

On the other hand, to maximize the half-duplex \dmt\ it is optimal
to choose $t=1/2$ whenever $m=n$. To see this, we compare
(\ref{eqn:C_1(t)}) with (\ref{eqn:C_2(t)}), and note that both
$K_{S,RD}^{\prime}\geq K_{S,D}$ and $K_{SR,D}\geq K_{S,D}$ for
$m=n$. Furthermore, for $m=n$ $d_{\mathcal{C}_S}'(r,t) =
d_{\mathcal{C}_D}'(r,1-t)$, and $d_{\mathcal{C}_S}'(r,t)$ is a
non-decreasing function in $t$. Therefore
$\min\{d_{\mathcal{C}_S}'(r,t), d_{\mathcal{C}_D}'(r,t)\}$ must
reach its maximum at $t=1/2$.

\subsection{Discussion}\label{subsec:MultiAntennaSingleRelayHDdiscussion}

When $(m,n,k) = (1,1,1)$, the best known half-duplex \dmt\ in the
literature is provided by the dynamic decode-and-forward (\ddf)
protocol~\cite{AzarianEGS05}. The \ddf\ protocol achieves
\begin{eqnarray*}
d_{DDF}(r) = \left \{ \begin{array}{lll}
  2(1-r) & \mbox{if} & 0 \leq r \leq \frac{1}{2}  \\
  \frac{1-r}{r} & \mbox{if} & \frac{1}{2} \leq r \leq 1 \\
\end{array}\right.,
\end{eqnarray*}
which does not meet the upper bound for $\frac{1}{2} \leq r \leq 1$,
as in this range, the relay does not have enough time to transmit
the high rate information it received. We would like to note that,
if the relay had all \csi, the \dmt\ of the \ddf\ protocol would not
improve. With this \csi\ the relay could at best perform beamforming
with the source; however, this only brings power gain, which does
not improve \dmt. It is also worth mentioning that when only relay
\csi\ is present, incremental \df\ \cite{LanemanTW02} would not
improve the \dmt\ performance of \df. Unless the source knows
whether the destination has received its message or not, it will
never be able to transmit new information to increase multiplexing
gains in incremental relaying.

\begin{figure}[t]
\centering
\includegraphics[width=3.4in]{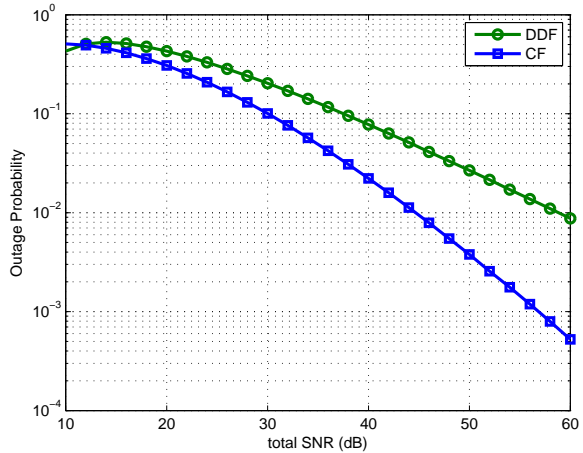}
\caption{Outage probability versus total \snr\ for the non-clustered
multiple-antenna, single half-duplex relay system,
$(m,n,k)=(2,2,1)$, $r = 1.5$, $t=0.5$.}
\label{fig:SRD_212_NonClustered_HD}
\end{figure}

In general it is hard to compute the \dmt\ of multiple antenna \ddf.
This is because the instantaneous mutual information \ddf\ achieves
in a multiple antenna relay channel is equal to
$I_{\mathcal{C}_D}(t)$ of (\ref{eqn:cutset2}) where $t$ is the
random time instant at which the relay does successful decoding.
Thus it is even harder to compute the \dmt\ for this case than for
fixed $t$. Moreover, we think that the multiple antenna \ddf\
performance will still be suboptimal. In
Section~\ref{sec:MultiAntennaSingleRelay}, we showed that for a
multiple antenna full-duplex relay system, the probability that the
relay cannot decode is dominant and the \df\ protocol becomes
suboptimal. Therefore, we do not expect any relay decoding based
protocol to achieve the \dmt\ upper bound in the multiple antenna
half-duplex system either. This conjecture is also demonstrated in
Fig.~\ref{fig:SRD_212_NonClustered_HD}, which shows the outage
probability versus total \snr\ for \ddf\ and \cf\ protocols for
$(m,n,k)=(2,2,1)$, $R^{(T)} = r \log (P_S+P_R)$, $r = 1.5$, $t=0.5$.
Source has twice the power relay has. The matrices $\bH_{SR}$,
$\bH_{SD}$ and $\bH_{RD}$ have i.i.d. complex Gaussian entries with
real and imaginary parts zero mean and variance $1/2$. We observe
that the diversity gain the \cf\ protocol achieves is approximately
0.90, whereas the \ddf\ protocol approximately achieves 0.47.

\section{The Multiple-Access Relay Channel}\label{sec:MARC}
\begin{figure}[h]
\begin{centering}
\includegraphics[width=2.8 in]{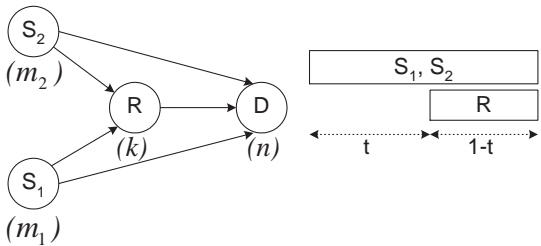}\\
\caption{System model for \textit{Problem 2}, the multiple access
relay channel. The sources, the destination and the relay have
$m_1$, $m_2$, $n$ and $k$ antennas
respectively.}\label{fig:MARCsystemmodel}
\end{centering}
\end{figure}

The most general network we introduced in
Section~\ref{sec:Introduction} includes the multiple access relay
channel (\marc) as a subproblem, (\textit{Problem 2}). The model for
MARC is shown in Fig.~\ref{fig:MARCsystemmodel}. Our emphasis is on
half-duplex \marc. As in Section~\ref{sec:MultiAntennaSingleRelayHD}
without loss of generality we consider a static, fixed state
protocol, where the relay listens for $t$ fraction of time,
transmits for $(1-t)$ fraction and sources transmit all the time.

For the half-duplex \marc\ we have
\begin{eqnarray*}
\bY_{R,1} & = & \bH_{S_1R} \bX_{S_1,1} + \bH_{S_2R} \bX_{S_2,1}+ \bZ_{R,1} \\
\bY_{D,1} & = & \bH_{S_1D} \bX_{S_1,1} + \bH_{S_2D} \bX_{S_2,1} +
\bZ_{D,1}
\end{eqnarray*} at state $q_1$ (when the relay listens) and at state $q_2$ (when the relay transmits),
the received signal at the destination is given as
\begin{eqnarray*}
\bY_{D,2} & = & \bH_{S_1D} \bX_{S_1,2} + \bH_{S_2D} \bX_{S_2,2}+
\bH_{RD} \bX_{R,2}+ \bZ_{D,2}.
\end{eqnarray*}
Here $\bX_{S_1,l}$, $\bX_{S_2,l}$ and $\bX_{R,l}$ are of size $m_1$,
$m_2$ and $k$ column vectors respectively and denote transmitted
signal vectors at node $S_1$, $S_2$ and $R$ at state $q_l$, for $l =
1,2$. Similarly $\bY_{R,l}$ and $\bY_{D,l}$ are the received signal
vectors of size $k$ and $n$. $\bH_{SD}$, $\bH_{{S_1}D}$,
$\bH_{{S_2}D}$, $\bH_{{S_1}R}$, $\bH_{{S_2}R}$, $\bH_{RD}$ are the
channel gain matrices of size $n \times m_1$, $n \times m_2$, $k
\times m_1$, $k \times m_2$, $n \times k$ respectively. The system
is non-clustered.

In this section we examine the \dmt\ for the \marc. We present our
results for the \marc\ with single antenna nodes to demonstrate the
basic idea.

The \dmt\ upper bound for the symmetric \marc\ occurs when both
users operate at the same multiplexing gain $r/2$, $\bar{r} =
(r/2,r/2)$, \footnote{This should not be confused with the notation
of~\cite{TseVZ04}, in which $r$ denotes the per user multiplexing
gain in case of symmetric users.} and is given in
\cite{AzarianEGS06} as
\begin{eqnarray}
 d_{MARC}(\bar{r})\leq \left \{ \begin{array}{lll}
  2-r & \mbox{if} & 0\leq r \leq \frac{1}{2} \\
  3(1-r) & \mbox{if} & \frac{1}{2} \leq r \leq 1\\
\end{array}\right. ,\label{eqn:marc_upperbound}
\end{eqnarray}which follows from cut-set upper bounds on the information rate.
Although this upper bound is a full-duplex \dmt\ bound, it is tight
enough for the half-duplex case when each node has a single antenna.
We see that this upper bound has the single user \dmt\ for $r \leq
\frac{1}{2}$, and has the relay channel \dmt\ with a two-antenna
source for high multiplexing gains. This is because for low
multiplexing gains, the typical outage event occurs when only one of
the users is in outage, and at high multiplexing gains, the typical
outage event occurs when both users are in outage, similar to
multiple antenna multiple-access channels \cite{TseVZ04}.

In Sections~\ref{sec:MultiAntennaSingleRelay} and
\ref{sec:MultiAntennaSingleRelayHD}, we have observed that \cf\ is
\dmt\ optimal for full-duplex and half-duplex multi-antenna relay
channels. This motivates us to study the performance of \cf\ in
\marc.

\begin{theorem}
For the single antenna, half-duplex \marc\ the \cf\ strategy
achieves the \dmt\ \[ d_{MARC, CF}(\bar{r}) = \left\{
\begin{array}{lll} 2(1-r) & \mbox{if} & 0 \leq r \leq \frac{2}{3} \\
1-\frac{r}{2}& \mbox{if} & \frac{2}{3} \leq r \leq \frac{4}{5} \\
3(1-r)& \mbox{if} & \frac{4}{5} \leq r \leq 1 \\
\end{array} \right. .\]
\label{thm:MARC_CF_DMT} This \dmt\ $d_{MARC, CF}(\bar{r})$ becomes
equal to the upper bound for $r \geq 4/5$.
\end{theorem}
\vspace{0.2 cm}
\begin{proof}
The proof is provided in Appendix~\ref{app:MARC_CF_DMT}.
\end{proof}

To achieve the above \dmt\ performance, two types of operation are
necessary. For low multiplexing gains, $0 \leq r \leq 2/3$, $S_1$
and $S_2$ utilize time sharing, and equally share the relay. Here
both $S_1$ and $S_2$ transmit for the half of the total time, for
$1/4$ of the whole time slot $R$ helps $S_1$ only, and in the last
quarter, $R$ helps $S_2$. Then we can directly apply the results
obtained in Section~\ref{sec:MultiAntennaSingleRelayHD}, which
results in the \dmt\ $2(1-r)$ in terms of the sum multiplexing gain.
For high multiplexing gains, $2/3 \leq r \leq 1$, both sources
transmit simultaneously. In this multiple access mode, for $4/5 \leq
r \leq 1$, both users being in outage is the dominant outage event,
the system becomes equivalent to the multiple antenna half-duplex
relay channel, and \cf\ achieves the \dmt\ upper bound.
\begin{figure}[t]
\begin{centering}
\includegraphics[width=3.4 in]{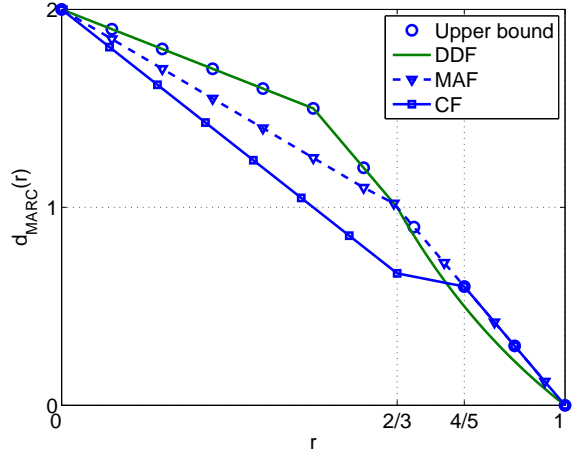}\\
\caption{\dmt\ for \marc. Each node has a single
antenna.}\label{fig:MARC_DMT}
\end{centering}
\end{figure}

For comparison the achievable \dmt\ with \ddf\ for \marc\
satisfies \cite{AzarianEGS06}:
\begin{eqnarray*}
d_{MARC, DDF}(r/2,r/2) \geq \left \{ \begin{array}{lll}
  2-r & \mbox{if} & 0 \leq r \leq \frac{1}{2} \\
  3(1-r) & \mbox{if} & \frac{1}{2} \leq r \leq \frac{2}{3} \\
  2 \frac{1-r}{r} & \mbox{if} & \frac{2}{3} \leq r \leq 1 \\
\end{array}\right. .
\end{eqnarray*}
We also compare our results with the \maf\ protocol for the \marc\
channel~\cite{ChenL06, ChenAL07} in Fig.~\ref{fig:MARC_DMT}. The
\maf\ performance is given as
\[ d_{MARC, MAF}(r/2,r/2) = \left \{ \begin{array}{ccc}
  2-3\frac{r}{2} & \mbox{if} & 0 \leq r \leq \frac{2}{3} \vspace{0.1 cm} \\
  3(1-r) & \mbox{if }&\frac{2}{3} \leq r \leq 1  \\
\end{array}\right. .\]

We observe that for low multiplexing gains, when single user outage
is dominant, it is optimal to decode the sources; however for high
multiplexing gains, compression works better. The \maf\ protocol is
also optimal for high multiplexing gains.

In Section~\ref{sec:MultiAntennaSingleRelay} we observed that for a
full-duplex relay channel, when terminals have multiple antennas,
\df\ becomes suboptimal, whereas \cf\ is not. Hence, we conjecture
that \ddf\ will not be able to sustain is optimality even in the low
multiplexing gain regime when the terminals have multiple antennas.
Moreover, it is not easy to extend the \maf\ protocol for multiple
antenna \marc. Even when we have one source, the \dmt\ for the
multiple antenna \naf\ protocol for the relay channel is not known,
only a lower bound exists~\cite{YangBelfiore07}. On the other hand,
for the multiple antenna case \cf\ will still be optimal whenever
decoding all sources together is the dominant error event. However,
for some antenna numbers $m_1$, $m_2$, and $n$, single-user behavior
will always dominate~\cite{TseVZ04}.

\section{Single Antenna Nodes, Multiple Relays}\label{sec:SingleAntenna}
In this section we examine \textit{Problem 3} and \textit{Problem 4}
to see how closely a cooperative system can mimic \mimo\ in terms of
\dmt. We first study a single source destination pair with 2 relays
(\textit{Problem 3}) in Section \ref{subsec:SingleAntennaTwoRelays},
then consider two sources and destinations (\textit{Problem 4}) in
Section \ref{subsec:SingleAntennaTwoSourceTwoDes}. In both cases,
each node has a single antenna. We also assume the nodes are
full-duplex so that we can observe the fundamental limitations a
relaying system introduces.
\begin{figure}[t]
\centering
\includegraphics[width=3.4 in]{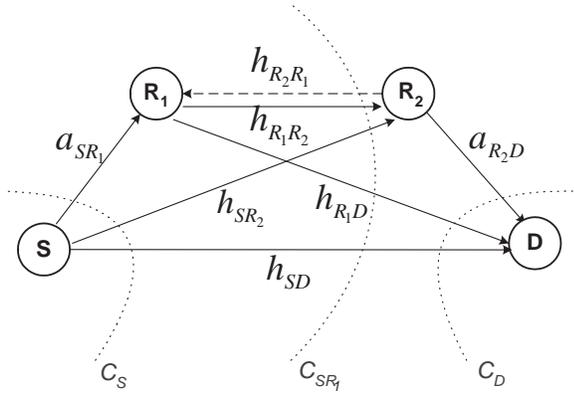}
\caption{System model for \textit{Problem 3}, the single
source-destination two relay system, each node has a single
antenna.} \label{fig:SingleAntennaTwoRelay}
\end{figure}

\subsection{Single Source-Destination, Two Relays}
\label{subsec:SingleAntennaTwoRelays} In this system there is a
single source-destination pair and two relays as shown in
Fig.~\ref{fig:SingleAntennaTwoRelay}. The channel is characterized
by
\begin{eqnarray}
Y_{R_1} &=& a_{SR_1}X_S+h_{R_2R_1}X_{R_2} +Z_{R_1} \label{eqn:SingAntTwoRelayR1receivedsignal}\\
Y_{R_2} &=& h_{SR_2}X_S+h_{R_1R_2}X_{R_1} +Z_{R_2} \label{eqn:SingAntTwoRelayR2receivedsignal}\\
Y_D &=& h_{SD}X_S+h_{R_1D}X_{R_1}+a_{R_2D}X_{R_2}+Z_D
\label{eqn:SingAntTwoRelayDreceivedsignal}
\end{eqnarray}
where $X_i$ and $Y_i$, $i= S,R_1,R_2,D$, are transmitted and
received signals at node $i$ respectively. The channel gains
$h_{ij}$, $i,j=S, R_1, R_2, D$, are independent, zero mean complex
Gaussian with variance $2\sigma^2$, where $\sigma^2$ is defined in
Section~\ref{sec:SystemModel}. As discussed in
Section~\ref{sec:SystemModel}, we assume the $R_2$ to $R_1$ link
$h_{R_2R_1}$, which is the dashed line in
Fig.~\ref{fig:SingleAntennaTwoRelay}, is present only if the system
is not clustered. If the system is not clustered, then the channel
gains $a_{ij}$ are also Rayleigh. On the other hand, if the system
is clustered, then $a_{SR_1}$ and $a_{R_2D}$ are equal to
$\sqrt{G_{SR_1}}$ and $\sqrt{G_{R_2D}}$ respectively, which are the
Gaussian channel gains. $Z_i$ denotes the \awgn\ noise, which is
independent at each receiver. The source, the first relay, $R_1$,
and the second relay, $R_2$, have power constraints $P_S$, $P_{R_1}$
and $P_{R_2}$ respectively. We assume the target data rate
$R^{(T)}=r \log \snr$. The following theorems summarize the main
results of this section.

\begin{theorem}\label{thm:SingleAntennaTwoRelayNonClus}The optimal \dmt\ for the non-clustered system of
Fig.~\ref{fig:SingleAntennaTwoRelay}, $d(r)$, is equal to \[ d(r) =
d_{13}(r).\] This optimal \dmt\ is achieved when both relays employ
\df\ strategy.
\end{theorem}
\vspace{0.2 cm}
\begin{proof} Please refer to
Appendices~\ref{app:SingleAntennaTwoRelayUB}
and~\ref{app:SingleAntennaTwoRelayNonClus} for the \dmt\ upper bound
and achievability results respectively.
\end{proof}
\begin{theorem}\label{thm:SingleAntennaTwoRelayClus}
The optimal \dmt\ for the clustered system of
Fig.~\ref{fig:SingleAntennaTwoRelay}, where $R_1$ is clustered with
the source and $R_2$ is clustered with the destination, $d(r)$ is
equal to \begin{eqnarray*} d(r) = \left \{\begin{array}{lll}
  d_{22}(r) & \mbox{if} & r \leq 1 \\
  0 & \mbox{if} & r > 1 \\
\end{array}\right.
.\label{eqn:SingleAntennaTwoRelayClus}\end{eqnarray*} The mixed
strategy, where $R_1$ does \df\ and $R_2$ does \cf\ achieves the
optimal \dmt.
\end{theorem}
\vspace{0.2 cm}
\begin{proof}Please refer to
Appendices~\ref{app:SingleAntennaTwoRelayUB}
and~\ref{app:SingleAntennaTwoRelayClus} for the \dmt\ upper bound
and achievability results respectively.
\end{proof}

\begin{figure}[t]
\centering
\includegraphics[width=3.4in]{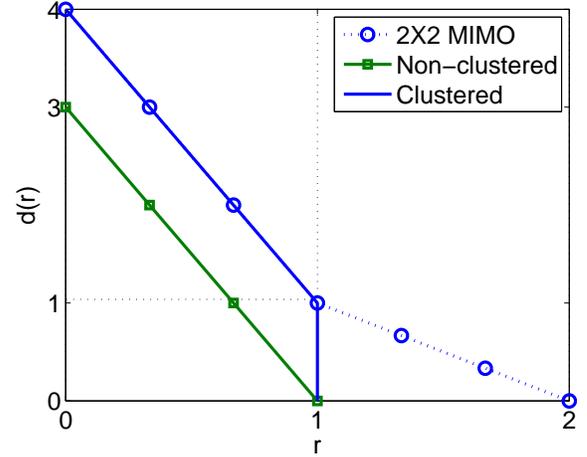}
\caption{\dmt\ for the single source-destination, two relay system,
each node has a single antenna.}
\label{fig:SingleAntennaTwoRelayDMT}
\end{figure}
Theorem~\ref{thm:SingleAntennaTwoRelayNonClus} says that if the
system is non-clustered it can at most have a transmit or a receive
antenna array \dmt\ behavior, but cannot act as a \mimo\ in terms of
\dmt. On the other hand, Theorem~\ref{thm:SingleAntennaTwoRelayClus}
confirms the fact that the multiplexing gain for the clustered
system is limited by 1. However, for all $r \leq 1$, the clustered
system can mimic a $2 \times 2$ \mimo, which means $d=1$ is
achievable at $r=1$.

\begin{figure}[t]
\centering
\includegraphics[width=3.4in]{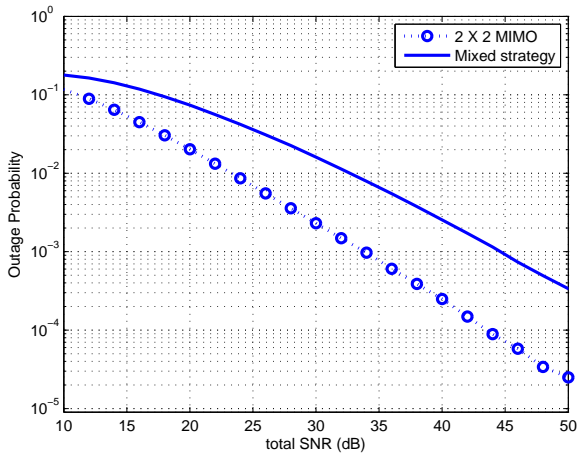}
\caption{Outage probability for the clustered single
source-destination, two relay system, each node has a single
antenna. $G_{SR_1} =G_{R_2D} = 10$, $r = 1$.}
\label{fig:SR1R2DclusteredMux1_FD}
\end{figure}

The \dmt\ performances for non-clustered and clustered systems as
well as $2 \times 2$ \mimo\ are illustrated in
Fig.~\ref{fig:SingleAntennaTwoRelayDMT}. We also display the outage
probability versus \snr\ for this clustered case in
Fig.~\ref{fig:SR1R2DclusteredMux1_FD} for $R^{(T)}=r \log
(P_S+P_{R_1}+P_{R_2})$, where $r=1$. We assumed $G_{SR_1} =G_{R_2D}
= 10$ and $P_S = P_{R_1} = 10 P_{R_2}$.  Also, $h_{ij}$, $i,j = S,
R_1, R_2, D$, are i.i.d. with $\sigma^2 = 1/2.$ For comparison, we
also show the outage probability of a $2 \times 2$ \mimo\ channel,
where the 2 transmit antennas share the total power equally and send
uncorrelated signals. The \mimo\ channel and the relay system have
the same total power constraint. We observe that as predicted, the
clustered relay network has the same diversity as the $2 \times 2$
\mimo\ and at $r=1$, $d=1$ is achievable.

We would like to note that for the clustered case \cf\ is essential
at $R_2$, and a strict decoding constraint at $R_2$ would limit the
system performance. If both relays do \df, $R_1$ will always be able
to decode for all multiplexing gains $0 \leq r \leq 1$, as the
$S-R_1$ channel can support rates up to $\log(1+G_{SR_1}P)$. Thus,
it is as if there is a two-antenna transmitter. However, $R_2$ may
or may not decode. Adapting
Appendix~\ref{app:SingleAntennaTwoRelayNonClus} to the clustered
case we can easily find the probability of outage at the destination
from (\ref{eqn:SingleAntennaTwoRelayNonClusteredDF}) as
$P(\mbox{outage at D}) \dot{=}  \snr^{-d_{31}(r)}$, which shows that
the decoding constraint at $R_2$ limits the system performance, the
system still operates as a transmit antenna array. Even though one
could improve upon this strategy by using the \df\ protocol
of~\cite{KramerGG04}, which allows the relays to process the signals
they hear from the source and the other relay jointly, this still
does not provide $2 \times 2$ \mimo\ behavior. In this case both the
destination and $R_2$ observe $2 \times 1$ \dmt, and $P(\mbox{outage
at D}) \dot{=} \snr^{-2d_{21}(r)}=\snr^{-4(1-r)}$, which is still
suboptimal as it cannot achieve the upper bound of
Theorem~\ref{thm:SingleAntennaTwoRelayClus}. Although the
destination can always understand $R_2$ reliably (because of the
clustering assumption), whenever both of them fail, the system is in
outage. However, for the receive cluster, \cf\ fits very well. If
the received signal at the destination has high power due to large
$h_{SD}$ and $h_{R_1D}$, then $R_2$ to destination channel has lower
capacity because in the decoding process $Y_D$ is treated as
interference. On the other hand, the correlation between the relay
and destination signals is higher and a coarse description $Y_{R_2}$
is enough to help the destination. However, if the side information
has low received power, the $R_2$ to destination channel has higher
capacity and $R_2$ can send the necessary finer information as the
correlation is less.

In~\cite[Theorem 4]{KramerGG04}, the authors prove an achievable
rate for a multiple relay system, in which some of the relays \df\
and the rest \cf. Furthermore, the relays that perform \cf\
partially decode the signals from the relays that perform \df.
Performing this partial decoding leads to higher achievable rates.
However, to achieve the \dmt\ upper bound, for both the
non-clustered and clustered cases, there is no need for partial
decoding and a simpler strategy is enough.

Note that same multiplexing limitations in
Theorem~\ref{thm:SingleAntennaTwoRelayClus} would occur when the
source has two antennas and a single antenna relay is clustered with
a single antenna destination or the symmetric case when the
destination has two antennas and a single antenna relay is clustered
with a single antenna source. These multiple antenna, single
source-destination, single relay cases were discussed in detail in
Section~\ref{sec:MultiAntennaSingleRelay}. In addition to these, we
investigate whether the multiplexing gain limitation is due to the
fact that there is only a single source-destination pair in the next
subsection.

\subsection{Two-Source Two-Destination Cooperative
System}\label{subsec:SingleAntennaTwoSourceTwoDes} 
\begin{figure}[t]
\begin{center}
\includegraphics[width=3.4 in]{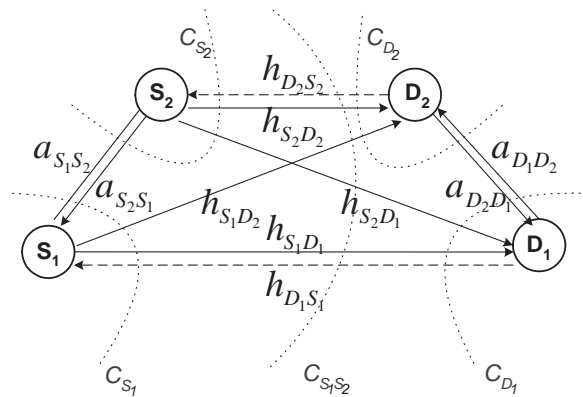}
 \caption{System model for \textit{Problem 4}, the two-source two-destination system, each node has a single antenna.}\label{fig:SingleAntennaTwoSourceTwoDes}
\end{center}
\end{figure}

We consider two
sources and two destinations, where sources cooperate in
transmission and destinations cooperate in reception
(\textit{Problem 4}). The system model is shown in
Fig.~\ref{fig:SingleAntennaTwoSourceTwoDes}. \textit{Problem 3}
studied in Section~\ref{subsec:SingleAntennaTwoRelays} would be a
special case of this, if one source has no information to send.

First we examine the multi-cast scenario, when both destinations are
required to decode both sources. This is analogous to \mimo\ systems
and represents the information transfer from a group of antennas to
another group of antennas. We define individual target data rates
$R^{(T_{S_1})} = r_1 \log \snr$ and $R^{(T_{S_2})} = r_2 \log \snr$,
with a sum target data rate of $R^{(T)}=r \log \snr$, $r = r_1+r_2$,
$\bar{r}=(r_1,r_2)$. Using the cut-set bounds in
Fig.~\ref{fig:SingleAntennaTwoSourceTwoDes} we have the following
corollary.
\begin{corollary}For a multi-cast, single antenna two-source
two-destination system, the system \dmt\ $d(\bar{r})$ is upper
bounded by
\[ d(\bar{r}) \leq d_{13}(r),\] if the system is non-clustered, and by
\[ d(\bar{r}) \leq d_{22}(r),~0 \leq r \leq 1,\] if the system is
clustered. Here $r = r_1+r_2$ is the sum multiplexing gain of the
system, and this upper bound is maximized for $\bar{r} = (r/2,r/2)$.
\end{corollary}\vspace{0.2 cm}
We omit the proof, which is very similar to the upper bound
calculation in Section~\ref{subsec:SingleAntennaTwoRelays}.

We observe that cooperative multicast is still limited in
multiplexing gains. We next study study the cooperative interference
channel, where $D_1$ is only required to decode $S_1$ and $D_2$ to
decode $S_2$. The cooperative interference channel imposes looser
decoding requirements on the destinations and potentially leads to
higher achievable rates. The next corollary shows that for the
clustered cooperative interference channel it is still not possible
to achieve multiplexing gains above $r=1$. Hence, we conclude the
multiplexing gain limitation is not due to having one
source-destination, but is due to the finite capacity links within
each cluster.

\begin{corollary}\label{thm:SingleAntennaCoopInf}
A single antenna two-source two-destination, clustered cooperative
interference channel has the best \dmt\ as
\[ \max_{r_1+r_2 = r}d(\bar{r}) = d_{22}(r),~0 \leq r \leq 1.\]
\end{corollary}
\vspace{0.2 cm}
\begin{proof}
We can show that $d(\bar{r}) \leq d_{22}(r)$ using the upper bound
of Lemma~\ref{Lemma:CutsetUpperBound}. The result in
\cite{HostMadsenN05} suggests that for cooperative interference
channel the total multiplexing gain can be at most 1. Thus we have
$d(\bar{r}) \leq d_{22}(r),~0 \leq r \leq 1.$

A simple achievable scheme assumes that ($S_1$, $D_1$) pair uses
($S_2$, $D_2$) pair as relays for half of the transmission period to
send $R^{(T_{S_1})} = r/2 \log \snr$. In the remaining half $S_2$
sends $R^{(T_{S_2})} = r/2 \log \snr$ to $D_2$ utilizing $S_1$ and
$D_1$ as relays. Note that equal distribution of rates gives the
best network diversity, since any other distribution of $r_1$ and
$r_2$ leads to a lower diversity for one of the streams. Then, for
each case, the problem reduces to the one discussed in
Section~\ref{subsec:SingleAntennaTwoRelays}. We can easily show that
this strategy meets the \dmt\ upper bound and that using such a time
division scheme is \dmt\ optimal for the cooperative interference
channel.
\end{proof}

For comparison, suppose there were two clustered single antenna
sources and a single two-antenna destination. This system can be a
\textit{virtual} \mimo, achieving the full $2 \times 2$ \mimo\ \dmt,
unlike the single-antenna two-source two-destination system
described above. In this case, when high diversity gains are needed,
the sources can cooperate, decode and forward each other's signals
using time division, and collectively act as a two-antenna
transmitter similar to the above argument. For high multiplexing
gains, they simply operate in the multiple access mode, i.e. each
source sends its own independent information stream, and thus can
attain all multiplexing gains up to 2~\cite{ZhengT03}. However, if
we had a two-antenna source and two clustered single antenna
destinations, the system would be multiplexing gain limited, as
\csi\ is not available at the transmitter \cite{JafarG05}. All these
examples emphasize the difference between transmit and receive
clusters, in addition to the effect of finite capacity links within
each cluster.

\section{Conclusion}\label{sec:Conclusion}
In this work we find the diversity-multiplexing tradeoff (\dmt) for
the following subproblems of a general multiple antenna network with
multiple sources, multiple destinations and multiple relays: 1) A
single source-destination system, with one relay, each node has
multiple antennas, 2) The multiple-access relay channel with
multiple sources, one destination and one relay, each node has
multiple antennas, 3) A single source-destination system with two
relays, each node has a single antenna, 4) A multiple
source-multiple destination system, each node has a single antenna.
For different configurations we consider the effect of half-duplex
or full-duplex behavior of the relay as well as clustering.

Firstly, we study a full-duplex multi-antenna relay system \dmt. We
examine the effect of clustering on both the \dmt\ upper bounds and
achievability results. We compare a single-antenna relay system with
a multiple-antenna relay system, when the source, the destination
and the relay have $m$, $n$ and $k$ antennas respectively, and
investigate the effects of increased degrees of freedom on the
relaying strategies decode-and-forward, and compress-and-forward. We
find that multi-antenna relay systems have fundamental differences
from their single-antenna counterparts. Increased degrees of freedom
affects the \dmt\ upper bounds and the performance of different
relaying strategies leading to some counterintuitive results.
Although the \df\ protocol is simple and effective to achieve the
\dmt\ upper bounds in single antenna relay systems, it can be
suboptimal for multi-antenna relay systems, even if the relay has
the same number of antennas as the source. On the other hand, the
\cf\ strategy is highly robust and achieves the \dmt\ upper bounds
for all multiplexing gain values for both clustered and
non-clustered networks. Clustering is essential for \df\ to achieve
the \dmt\ upper bound for low multiplexing gains, but does not help
in the high multiplexing gain region. What's more, it has an adverse
effect on both the upper bound and the \df\ achievable \dmt\ if the
relay has multiple antennas due to decreased degrees of freedom in
the source-relay channel.

We extend the above full-duplex results obtained for the multiple
antenna relay channel to the half-duplex relay as well. We show that
for the multiple-antenna half-duplex relay channel the \cf\ protocol
achieves the \dmt\ upper bound. Although it is hard to find the
\dmt\ upper bound explicitly for arbitrary $m$, $n$ and $k$, we have
solutions for special cases. We show that the half-duplex \dmt\
bound is tighter than the full-duplex bound in general, and \cf\ is
\dmt\ optimal for any $m$, $n$ and $k$. We also argue that the
dynamic decode-and-forward protocol or any decoding based protocol
would be suboptimal in the multiple antenna half-duplex relay
channel as they are suboptimal in the full-duplex case.

We next investigate the multiple-access relay channel. In \marc,
\cf\ achieves the upper bound for high multiplexing gains, when both
users being in outage is the dominant outage event.

Finally, we compare wireless relay and cooperative networks with a
physical multi-input multi-output system. We show that despite the
common belief that the relay or cooperative systems can be
\textit{virtual} \mimo\ systems, this is not possible for all
multiplexing gains. Both for relay and cooperative systems, even if
the nodes are clustered, the finite capacity link between nodes in
the source cluster and the finite capacity link between the nodes in
the destination cluster are bottlenecks and limit the multiplexing
gain of the system. Cooperative interference channels are also
limited the same way. It is straightforward to extend our results
for a single source-destination pair with multiple relays and for
cooperative systems with N sources and N destinations with each
destination decoding all sources.

Overall, our results indicate the importance of soft information
transmission in relay networks, as in \cf, and suggest that protocol
design taking into account node locations, antenna configurations
and transmission/reception constraints are essential to harvest
diversity and multiplexing gains in cooperative systems.

\appendices
\section{Proof of
Lemma~\ref{Lemma:CutsetUpperBound}}\label{app:LemmaCutsetBound} By
Proposition~\ref{thm:CutsetBound}, the information rates $R^{(kl)}$
from node $k$ to node $l$ in the network satisfy
\begin{eqnarray*} \sum_{k \in \mathcal{C}_i, l \in \mathcal{C}_i^c }R^{(kl)}
&\leq& I_{\mathcal{C}_i}
\end{eqnarray*}
for some $p(x_1,x_2,...,x_m)$. Also, we can easily observe that
$\mathcal{E}_{N}$ is implied by the event
\[\sum_{k \in \mathcal{C}_i, l \in \mathcal{C}_i^c }R^{(kl)} <
R^{(T_{\mathcal{C}_i})}. \] Then for any coding scheme with rates
$R^{(kl)}$, we can write
\begin{eqnarray*}
P\left(\mathcal{E}_{N} \right)
 &\geq & P\left(\sum_{k \in \mathcal{C}_i,
l \in \mathcal{C}_i^c }R^{(kl)} < R^{(T_{\mathcal{C}_i})}\right) \\
&\geq & P( I_{\mathcal{C}_i} < R^{(T_{\mathcal{C}_i})}) \nonumber\\
&\geq& \min_{p(x_1,x_2,...,x_m)} P(  I_{\mathcal{C}_i} <
R^{(T_{\mathcal{C}_i})}). \label{eqn:lemmaAux}
\end{eqnarray*}
The above statement holds for all coding schemes with rates
$R^{(kl)}$; thus, it is also true for the one that minimizes the
left hand side. Then we have
\begin{eqnarray}
\min_{\mbox{\tiny{all coding schemes}}} P\left(\mathcal{E}_N\right)
 \geq \min_{p(x_1,x_2,...,x_m)} P( I_{\mathcal{C}_i}
< R^{(T_{\mathcal{C}_i})}) \label{eqn:app11}
\end{eqnarray}
The right hand side is the minimum outage probability for cut-set
$\mathcal{C}_i$, and by the definition in the lemma
\begin{eqnarray} \min_{p(x_1,x_2,...,x_m)} P( I_{\mathcal{C}_i} <
R^{(T_{\mathcal{C}_i})}) \dot{=}
\snr^{-d_{\mathcal{C}_i}(r^{(\mathcal{C}_i)})}.\label{eqn:app12}\end{eqnarray}
Using the definition of maximum network diversity, we have
\begin{eqnarray} \min_{\mbox{\tiny{all coding schemes}}}
P\left(\mathcal{E}_N\right)  \dot{=}
\snr^{-d(\bar{r})}.\label{eqn:app13} \end{eqnarray} Substituting
(\ref{eqn:app12}) and (\ref{eqn:app13}) into (\ref{eqn:app11}) leads
to
\[ d(\bar{r}) \leq d_{\mathcal{C}_i}(r^{(\mathcal{C}_i)}).\]
Since this is true for all the cut-sets, we have
\[ d(\bar{r}) \leq \min_i d_{\mathcal{C}_i}(r^{(\mathcal{C}_i)}).\]
We conclude that the maximum network diversity order $d(\bar{r})$ is
upper bounded by the maximum diversity order of each
$I_{\mathcal{C}_i}$.

\section{Proof of Theorem~\ref{thm:MultiAntennaSingleRelayNonClusDF}}\label{app:MultiAntennaSingleRelayNonClusDF}

In the \df\ protocol, the source and the relay employ block Markov
superposition coding~\cite{Cover, KramerGG04} and the destination
does backward decoding~\cite{KramerGG04,ZengKBuzo89,WillemsvM85}.
For achievability, we constrain the relay to decode the source
signal reliably. If based on its received \snr\ the relay cannot
decode, then it remains silent (or sends a default signal). We
assume this is known at the destination, which can be communicated
at a negligible cost. Since fading is constant for all $B$ blocks,
this has to be communicated only once.

The relay decodes if the instantaneous mutual information satisfies
\[ R^{(DF)} \leq I(\bX_S;\bY_{R}|\bX_{R}).\] If the relay can
decode, the mutual information at the destination is
$I(\bX_S\bX_R;\bY_D)$, otherwise it is $I(\bX_S;\bY_D|\bX_{R})$. We
choose $\bX_S$ and $\bX_R$ independently as complex Gaussian with
zero mean with covariance matrices $Q_S = \mathbf{I}_mP_S/m$, and
$Q_R = \mathbf{I}_kP_R/k$ respectively. Then we can write
\begin{eqnarray*}
I(\bX_S;\bY_{R}|\bX_{R}) & = & \log L_{S,R} \label{eqn:DF1}\\
I(\bX_S;\bY_D|\bX_{R}) & = & \log L_{S,D}\label{eqn:DF3} \\
I(\bX_S\bX_R;\bY_D) &= &\log L_{SR,D}, \label{eqn:DF2}
\end{eqnarray*}
where \[ L_{S,R} =\left| \mathbf{I}_k + \bH_{SR}\bH_{SR}^\dag
\frac{P_S}{m} \right|, \] and $L_{S,D}$ and $L_{SR,D}$ are defined
in (\ref{eqn:L_S,D}) and (\ref{eqn:L_SR,D}) respectively.

We calculate the probability of outage as
\begin{eqnarray}
\lefteqn{P(\mbox{outage at D}) }\nonumber\\
&=&P(\mbox{outage}|\mbox{relay decodes})P(\mbox{relay
decodes})\nonumber \\
&&{+}\: P(\mbox{outage}|\mbox{relay cannot decode})\nonumber \\
&&{.}\: P(\mbox{relay
cannot decode})\label{eqn:MultiAntennaSingleRelayDFout}\\
& = &P(L_{SR,D}< \snr^r) P(L_{S,R} > \snr^r) \nonumber \\
&&{+}\:P(L_{S,D}< \snr^r) P(L_{S,R} < \snr^r) \nonumber \\
&\dot{=}&\left\{
\begin{array}{l}
  {\snr^{-d_{(m+k)n}(r)}}+{\snr^{-d_{mn}(r)}}{\snr^{-d_{mk}(r)}} \\
~~~~~~~~~~~~~~~~~~~~~~~~~ \mbox{if~}  0\leq r\leq \min \{m,n,k\}\\
  {\snr^{-d_{mn}(r)}} \\
  ~~~~~~~~~~~~~ \mbox{if~}  \min\{m,n,k\} < r\leq \min \{m,n\}
\end{array} \right. \nonumber 
\end{eqnarray}
for which we used the fact that $P(L_{S,R} > \snr^r) \dot{=} 1$ for
$0 \leq r \leq \min \{m,n,k\}$, and for $\min\{m,n,k\}< r \leq \min
\{m,n\}$, $ P(L_{S,R} > \snr^r) \dot{=} 0$ and $P(L_{S,R} <
\snr^r)\dot{=} 1$. Hence we can write the \dmt\ for \df\ as in
(\ref{eqn:MultiAntennaNonClusDFdmt}). Note that any other choice of
$Q_S$, $Q_R$ and $Q = \mathrm{Cov}(\bX_S, \bX_R)$ would not improve
this result. This is because for any $Q_S$, $Q_R$ and $Q$, due to
(\ref{eqn:covarianceremoval}), the mutual information expressions
have the upper bounds
\begin{eqnarray*}
I(\bX_S;\bY_{R}|\bX_{R}) & \leq & \log \left| \mathbf{I}_k +
\bH_{SR}\bH_{SR}^\dag P_S \right| \label{eqn:DF1_UB}\\
I(\bX_S;\bY_D|\bX_{R}) & \leq & \log \left| \mathbf{I}_{n}+
\bH_{SD}\bH_{SD}^\dag P_S \right|\label{eqn:DF3_UB}
\end{eqnarray*}and
\begin{eqnarray*}
I(\bX_S \bX_R; \bY_D) & \leq & \log K_{SR,D} \label{eqn:DF2_UB}
\end{eqnarray*}
where $K_{SR,D}$ is defined in (\ref{eqn:K_SR,D}). A \dmt\
calculation using these upper bounds would result in the same \dmt\
as in (\ref{eqn:MultiAntennaNonClusDFdmt}).

\section{Proof of Theorem~\ref{thm:MultiAntennaSingleRelayClusUB}}\label{app:MultiAntennaSingleRelayClusteredUB}
First, the mutual information for cut-set $\mathcal{C}_D$ is the
same as the non-clustered case
of~(\ref{eqn:MultiAntennaSingleRelayCutsetD}), and the \dmt\ upper
bound for this cut-set is $d_{(m+k)n}(r)$. For cut-set
$\mathcal{C}_S$ we need to find the \dmt\ for the channel
\[ \mathbf{Y} = {\bH_{S,RD}}\mathbf{X}+\mathbf{Z},\]
where $\bH_{S,RD}$ is defined in (\ref{eqn:H_S_RD}), $\bH_{SR}$ is
an $k \times m$ matrix with all entries equal to $\sqrt{G}$, and
$\bH_{SD}$ is $n \times m$, with complex Gaussian entries $h_{ij}$,
$i=1,...,n$, $j = 1,2$. The channel input $\bX$ is $m \times 1$ and
has the total transmit power constraint $P_S$. We assume $P_S =
\snr$ for notational simplicity in this appendix. The channel
output, $\bY$, and the complex Gaussian noise at the output, $\bZ$,
are $(n+k)\times 1$.

For $m=1$, the \dmt\ is easily calculated as $d_{\mathcal{C}_S}(r) =
\infty$, $0 \leq r \leq 1$, as $I_{\mathcal{C}_S} > \log (1+k G
P_S)$.

For $m=2$, the instantaneous mutual information for a given channel
gain matrix $\bH_{S,RD}$ is then
\begin{eqnarray*}
I\left(\mathbf{X};\mathbf{Y}\right) &=& \log \left| \mathbf{I}_2 +
\snr\bH_{S,RD}^{\dag}\bH_{S,RD} \right|.
\end{eqnarray*}
Note that
\begin{eqnarray*}
\bH_{S,RD}^{\dag}\bH_{S,RD}& =& \left[  \bH_{SR}^\dag
\bH_{SR} +\bH_{SD}^\dag \bH_{SD} \right] \\
& =& \left[  \left(%
\begin{array}{cc}
  k G & k G \\
  k G & k G \\
\end{array}
\right) +\bH_{SD}^\dag \bH_{SD} \right],
\end{eqnarray*}
which means having $k$ relay antennas only increases the Gaussian
channel gain in between the source and the relay antennas by a
constant factor. Therefore, without loss of generality we can assume
$k=1$. For $k=1$,
\begin{eqnarray*}
{I(\mathbf{X};\mathbf{Y})} &=& \log \left( 1+2G \snr+
\Sigma_{i=1}^{n}\Sigma_{j=1}^{2}|h_{ij}|^2\snr +
 \right. \\
&& {+}\: \left.\Sigma_{i=1}^{n}|h_{i1}-h_{i2}|^2G \snr \right. \\
&& {+}\: \left.
\Sigma_{i=1}^{n}\Sigma_{j=i+1}^{n}|h_{i1}h_{j2}-h_{i2}h_{j1}|^2
\snr^2\right).
\end{eqnarray*}

There is no outage for multiplexing gain $r\leq 1$ as
$I(\mathbf{X};\mathbf{Y}) \geq\log (1+2G \snr)$. For $1 < r \leq 2$,
we can lower bound the outage probability as
\begin{eqnarray*}
P(\mbox{outage}) &=& P(\mbox{outage}|\mathcal{E})P(\mathcal{E}) +
P(\mbox{outage}|\mathcal{E}^c)P(\mathcal{E}^c) \\
& \geq & P(\mbox{outage}|\mathcal{E})P(\mathcal{E}),
\end{eqnarray*} where
\begin{eqnarray*}{\mathcal{E}} &=&\left\{ |\Re\{h_{i1}\}-\Re\{h_{i2}\}|<\epsilon,
|\Im\{h_{i1}\}-\Im\{h_{i2}\}|<\epsilon : \right. \\
&& \left. i=1,...,n \right\}
\end{eqnarray*}
with $\epsilon=1/\sqrt{\snr^{2-r}}$. When $\mathcal{E}$ holds,
\begin{eqnarray*} I(\bX;\bY) &=& \log \left( 1+2G \snr+
\Sigma_{i=1}^{n}\Sigma_{j=1}^{2}|h_{ij}|^2 \snr \right.\\
&&{+}\:  \left. 2 \epsilon^2 n G \snr \right. \\
&&{+}\: \left. 2 \epsilon^2
\Sigma_{i=1}^{n}\Sigma_{j=i+1}^{n}|h_{i1}-h_{j1}|^2 \snr^2 \right).
\end{eqnarray*}
Then for a target data rate $R^{(T)}=r \log \snr$, we have
\begin{eqnarray*}
{P(\mbox{outage}| \mathcal{E})} & =& P \left(
\Sigma_{i=1}^{n}\Sigma_{j=1}^{2}|h_{ij}|^2 \frac{1}{\snr^{r-1}}\right.\\
&& {+}\: \left.2 \Sigma_{i=1}^{n}\Sigma_{j=i+1}^{n}|h_{i1}-h_{j1}|^2
< f(\snr)
 \right),
\end{eqnarray*}
where \[f(\snr) =  \frac{\snr^r -1 - 2G \snr -2nG
\snr^{r-1}}{\snr^r}.
\]
Then for $ \snr > 1$, $1/\snr^{r-1}<1$, therefore, we can further
lower bound $P(\mbox{outage}| \mathcal{E})$ as
\begin{eqnarray*}
P(\mbox{outage}| \mathcal{E}) &\geq& P \left(
\Sigma_{i=1}^{n}\Sigma_{j=1}^{2}|h_{ij}|^2 \right. \\
&&{+}\: \left. 2 \Sigma_{i=1}^{n}\Sigma_{j=i+1}^{n}|h_{i1}-h_{j1}|^2
< f(\snr)
 \right).
\end{eqnarray*}
As $f(\snr) \dot{=} 1$, then $P(\mbox{outage}|\mathcal{E})\dot{=}1$.
On the other hand, as the real and imaginary parts of all random
variables are i.i.d. we have
\begin{eqnarray*}
P(\mathcal{E})= ( P( |\Re\{h_{i1}\}-\Re\{h_{i2}\}|<\epsilon ))^{2n},
\end{eqnarray*}
and
\begin{eqnarray*}
\lefteqn{P( |\Re\{h_{i1}\}-\Re\{h_{i2}\}|<\epsilon )}\\
 & = &
\int_{-\infty}^{\infty}
\int_{x-\epsilon}^{x+\epsilon} \frac{1}{\pi}e^{-y^2-x^2}dy dx \\
& = & \int_{-\infty}^{\infty} \int_{-\epsilon}^{\epsilon}
\frac{1}{\pi}e^{-(t+x)^2}e^{-x^2}dt dx \\
& = & \int_{-\epsilon}^{\epsilon} \int_{-\infty}^{\infty}
\frac{1}{\pi}e^{-2x^2-2tx}e^{-t^2}dx dt \\
& = & \int_{-\epsilon}^{\epsilon} \frac{1}{\sqrt{2\pi}}e^{-t^2/2} dt \\
& = &\mathrm{erf}\left( \frac{\epsilon}{\sqrt{2}}\right).
\end{eqnarray*}

The error function has the Maclaurin series expansion of
\[ \mathrm{erf}(x) = \frac{2}{\sqrt{\pi}}\left(x - \frac{1}{3}x^3 + \frac{1}{10}x^5- \frac{1}{42}x^7 + ...\right),\]
which makes $\mathrm{erf}( \epsilon/\sqrt{2}) \dot{=} \epsilon$ at
high \snr. Then $ P(\mathcal{E}) \dot{=} \epsilon^{2n} =
\snr^{-2n+nr}, $ and we have \[P(\mathrm{outage}) \dot{\geq}
\snr^{-2n+nr}.\]

On the other hand,
\begin{eqnarray*}
\lefteqn{P(\mathrm{outage})} \\
& \leq & P\left( \log \left| \mathbf{I}_2 + \snr
\tilde{\bH}_{S,RD}^\dag \tilde{\bH}_{S,RD}^\dag \right| < r \log
\snr\right) \\
& \dot{=} & \snr^{-2n+nr},
\end{eqnarray*}
where $\tilde{\bH}_{S,RD}^\dag = \left[ \tilde{\bH}_{SR}^\dag
~\bH_{SD}^\dag \right]$, and $\tilde{\bH}_{SR}$ is an $1 \times m$
matrix with i.i.d. complex Gaussian entries.

Thus, for $1<r \leq 2$, $d_{\mathcal{C}_S}(r)$ is equal to the \dmt\
of a $2 \times (n+1)$ system, $d_{2(n+1)}(r)=n(2-r)$, and overall we
have the $d(r)$ expression stated in
Theorem~\ref{thm:MultiAntennaSingleRelayClusUB} for $m=1,2$ and
arbitrary $n$ and $k$.

\section{Proof of
Lemma~\ref{Lemma:DMTboundHD}}\label{app:LemmaDMTboundHD} In random
state half-duplex relay systems, the system state can also be viewed
as a channel input. Thus, we need to optimize over all joint
distributions $p(x_S,x_R,q)$. Using
Proposition~\ref{thm:CutsetBound} we have
\[ R^{(SD)} \leq \min_i I_{\mathcal{C}_i},\]
$i = S,D$, for some $p(x_S,x_R,q)$, where $R^{(SD)}$ is the
information rate from $S$ to $D$. Then for a target data rate
$R^{(T)}= r \log \snr$ we have
\begin{eqnarray*} \lefteqn{ \min_{\mbox{all coding schemes}}
P\left(R^{(SD)} < R^{(T)}\right) }\\& \geq& \min_{p(x_S,x_R,q)}
\max_{i }  P( I_{\mathcal{C}_i} < R^{(T)})\\  &\geq& \min_{p(q)}
\max_{i } \min_{p(p(x_S,x_R|q))} P( I_{\mathcal{C}_i} < R^{(T)}).
\end{eqnarray*}
Then using (\ref{eqn:div_HD}) we can write
\begin{eqnarray*}d({r}) \leq \max_{p(q)} \min_i
d_{\mathcal{C}_i}(r,p(q)).\label{eqn:dr}\end{eqnarray*}

For the multiple antenna, half-duplex relay channel we have
\begin{eqnarray*}
I_{\mathcal{C}_S} & = & I(X_S, Q;Y_R, Y_D|X_R)\nonumber \\
& = & I(X_S; Y_R, Y_D|X_R, Q) + I(Q;Y_R,Y_D|X_R)\nonumber \\
& \leq & I(X_S; Y_R, Y_D|X_R, Q) + 1 \label{eqn:1}
\end{eqnarray*}
where the last inequality follows because $Q$ is a binary random
variable. Similarly,
\begin{eqnarray*}
I_{\mathcal{C}_D} & = & I(X_S,X_R,Q; Y_D) \nonumber\\
& = & I(X_S, X_R; Y_D|Q) + I(Q;Y_D)\nonumber \\
& \leq & I(X_S, X_R; Y_D|Q) + 1. \label{eqn:2}
\end{eqnarray*}
The above two bounds show that random state protocols can at most
send one extra bit of information, which does not play a role at
high \snr. Thus, fixed and random state protocols have the same
\dmt\ upper bound.

\section{Proof of
Theorem~\ref{thm:MultiAntennaSingleRelayHD_CF}}\label{app:MultiAntennaSingleRelayHD_CF}
To illustrate that \cf\ achieves the \dmt\ in
Theorem~\ref{thm:MultiAntennaSingleRelayHD_CF}, we follow the \cf\
protocol of Section~\ref{sec:MultiAntennaSingleRelay}. In the static
half-duplex case the relay listens to the source only for $t$
fraction of time with $0 \leq t \leq 1$. The Wyner-Ziv type
compression rate is such that the compressed signal at the relay can
reach the destination error-free in the remaining $(1-t)$ fraction
of time, in which the relay transmits. Then, for a fixed~$t$ the
instantaneous mutual information at the destination is
\begin{eqnarray*}
R^{(CF)} = t I(\bX_S;\hat{\bY}_R\bY_D|q_1) +
(1-t)I(\bX_S;\bY_D|\bX_R,q_2) \label{eqn:RC_CF}
\end{eqnarray*}
subject to
\begin{eqnarray}
t I(\hat{\bY}_R;\bY_R|\bY_D, q_1)\leq (1-t)I(\bX_R;\bY_D|q_2).
\label{eqn:CompConstrHD}
\end{eqnarray}
Note that the above equations incorporate the half-duplex constraint
into (\ref{eqn:CompConstr}) and
(\ref{eqn:MultiAntennaSingleRelayCFrate}). The source and relay
input distributions are independent, $\hat{\bY}_R$ is the auxiliary
random vector which denotes the compressed signal at the relay and
depends on $\bY_R$ and $\bX_R$. More information on \cf\ can also be
found in \cite{HostMadsenZ05,LaiLiuEG05,LaiLiuEG06} for the
half-duplex case for single antenna nodes.

We consider $\bX_S$ and $\bX_R$ are i.i.d. complex Gaussian with
zero mean and covariance matrices $\mathbf{I}_mP_S/m$,
$\mathbf{I}_kP_R/k$, $\hat{\bY}_R=\bY_{R,1}+\hat{\bZ}_R$, and
$\hat{\bZ}_R$ is a vector with i.i.d. complex Gaussian entries with
zero mean and variance $\hat{N}_R$ that is independent from all
other random variables. Using the definitions of $L_{S,D}$,
$L_{SR,D}$, $L_{S,RD}$, and $L_{S,RD}^{\prime}$ (\ref{eqn:L_S,D}),
(\ref{eqn:L_SR,D}), (\ref{eqn:L_S,RD}) and (\ref{eqn:L_S,RD'}) we
have
\begin{eqnarray*}
 I(\hat{\bY}_R;\bY_R|\bY_D,q_1) &=&
\log\frac{L_{S,RD}}{L_{S,D}\hat{N}_R^k},\\ I(\bX_R;\bY_D|q_2) &=&
\log \frac{L_{SR,D}}{L_{S,D}}.
\end{eqnarray*}
Thus using (\ref{eqn:CompConstrHD}) we can choose the compression
noise variance $\hat{N}_R$ to satisfy
\begin{equation*}
\hat{N}_R=\sqrt[k]{\frac{L_{S,RD}}{U}},
 \mbox{~with~}
 U = L_{S,D}
\left(\frac{L_{SR,D}}{L_{S,D}}\right)^{(\frac{1-t}{t})},\label{eqn:RelayChannelHalfDuplexCFNoise}\end{equation*}
and (\ref{eqn:RC_CF}) becomes \begin{eqnarray} R^{(CF)} = t
\log\frac{L_{S,RD}}{\left({\hat{N}_R}+1\right)^k} + (1-t)\log
L_{S,D}. \label{eqn:RC_HD_CF}
\end{eqnarray}

To prove the \dmt\ of (\ref{eqn:RC_HD_CF}) we follow steps similar
to
(\ref{eqn:MultiAntennaSingleRelayNonClusPout1})-(\ref{eqn:MultiAntennaSingleRelayNonClusPout2}).
Then we have
\begin{eqnarray}
\lefteqn{P(\mbox{outage at D})}\nonumber \\
&=&P\left(R^{(CF)}<r\log \snr\right) \nonumber \\
&=& P \left( {L_{S,RD}^{\prime}}^tL_{S,D}^{(1-t)} < 2^{kt}\snr^{r}
\right) \nonumber \\ &&{+}\: P \left( L_{SR,D}^{(1-t)}L_{S,D}^{t} <
2^{kt}\snr^{r}
\right)\nonumber \\
 &\overset{(a)}{\leq} & P \left(
{L_{S,RD}^{\prime}}^tL_{S,D}^{(1-t)} < 2^k\snr^{r} \right)\nonumber
\\ &&{+}\:  P \left( L_{SR,D}^{(1-t)}L_{S,D}^{t} < 2^k\snr^{r} \right)
\label{eqn:CFproof}\\
 & = & P \left( t \log{L_{S,RD}^{\prime}}\right. \nonumber \\ &&{+}\:
\left.(1-t)\log L_{S,D} < r
\log \sqrt[r]{2^k}\snr \right) \nonumber\\
&&{+}\: P \left((1-t)\log L_{SR,D}\nonumber \right.\\ &&{+}\: \left. t \log L_{S,D} < r \log \sqrt[r]{2^k}\snr \right)  \label{eqn:sss}\\
& \overset{(b)}{\dot{=}} &\snr^{-d_{\mathcal{C}_S}'(r,t)}+
\snr^{-d_{\mathcal{C}_D}'(r,t)} \nonumber
\\
& \dot{=} & \snr^{-\min\{d_{\mathcal{C}_S}'(r,t),
d_{\mathcal{C}_D}'(r,t)\}}, \nonumber
\end{eqnarray}
where $(a)$ is because for any fixed $0\leq t \leq 1$,  $2^k>
2^{kt}$. For $(b)$ we have used the fact that $L_{S,RD}^{\prime}$
and $K_{S,RD}^{\prime}$, $L_{S,D}$ and $K_{S,D}$, and $L_{SR,D}$ and
$K_{SR,D}$ are of the same form except for power scaling and hence
result in the same \dmt. As a result if $P(\mbox{outage at D})
\dot{=} \snr^{-d(r,t)}$, then $d(r,t) \geq
\min\{d_{\mathcal{C}_S}'(r,t), d_{\mathcal{C}_D}'(r,t)\}$. As the
achievable \dmt\ cannot be larger than the upper bound, we conclude
that \cf\ achieves the bound in (\ref{eqn:RC_DMT_upperbound}) for
any $t$. Thus it also achieves the best upper bound of
(\ref{eqn:RC_DMT_upperbound_maximized}).

If the relay is dynamic, \cf\ can also behave dynamically and $t$
will be a function of \csi\ available at the relay. For dynamic \cf\
we can still upper bound the probability of outage at the
destination with (\ref{eqn:sss}), which is equivalent to the \dmt\
upper bound for dynamic protocols at high \snr. Hence, dynamic \cf\
achieves the dynamic half-duplex \dmt\ upper bound.

\section{Proof of Theorem~\ref{thm:HalfDuplexDMTBound}}\label{app:HalfDuplexDMTBound} In this appendix we prove
Theorem~\ref{thm:HalfDuplexDMTBound}. For $m =1$, (\ref{eqn:C_1(t)})
can be written as
\[I_{\mathcal{C}_S}'(t) = t\log K_{S,RD}^{\prime} + (1-t)\log K_{S,D}, \] with
\begin{eqnarray*}
K_{S,RD}^{\prime} &= &\log \left( 1+
\sum_{i=1}^{n+k}x_i\snr\right) \\
K_{S,D}& =& \log \left( 1+ \sum_{i=1}^{n}x_i\snr\right)
\end{eqnarray*}
where $x_i$ are independent exponentially distributed random
variables with parameter 1, that denote the fading power from source
antenna to receive antenna $i$ at the destination or at the relay
respectively.

Let $x_i = \snr^{-\alpha_i}$, $i = 1,...,n+k$. Then $\alpha_i$ are
i.i.d. with probability density function
\[f_{\alpha_i} (\alpha_i) =
\log(\snr)\snr^{-\alpha_i}\mathrm{exp}(-\snr^{-\alpha_i}).\]

Let $\mathcal{A}$ denote the outage event for a target data rate
$R^{(T)}=r\log\snr$. Then probability of outage is
\begin{eqnarray*}
 \lefteqn{P(\mathcal{A})} \\
   &= &P(I_{\mathcal{C}_S}'(t)<r \log\snr) \\
& = & \int_{\mathcal{A}}f_{\boldsymbol{\alpha}}(\boldsymbol{\alpha})d\boldsymbol{\alpha} \\
& = &
\int_{\mathcal{A}}(\log\snr)^{n+k}\snr^{-\Sigma \alpha_i}\exp(-\Sigma \snr^{-\alpha_i})d\boldsymbol{\alpha}\\
& \overset{(a)}{\dot{=}} & \int_{\mathcal{A}\bigcap
\mathbb{R}^{(n+k)+}}\snr^{-\Sigma \alpha_i}d\boldsymbol{\alpha} \\
& \overset{(b)}{\dot{=}} & \int_{\mathcal{\tilde{A}}\bigcap
\mathbb{R}^{(n+k)+}}\snr^{-\Sigma \alpha_i} d\boldsymbol{\alpha} \\
&\overset{(c)}{\dot{=}}&\mathrm{SNR}^{-G^*}
\end{eqnarray*}
where $\mathbb{R}^{(n+k)+}$ is the set of real $(n+k)$-vectors with
nonnegative elements. The outage event $\mathcal{\tilde{A}}$ is
defined as
\begin{eqnarray*} \mathcal{\tilde{A}}& =& \left \{t\max\{
0,1-\alpha_1, 1-\alpha_{n+1}\right. \} \\
&&{+}\:\left. (1-t) \max\{0,1-\alpha_1\} < r\right\}
\end{eqnarray*} where without loss of generality we assume $\alpha_1 = \min \{\alpha_1,
...\alpha_n\}$ and $\alpha_{n+1} = \min\{\alpha_{n+1}, ...,
\alpha_{n+k}\}, $
 and $G^*$ is given as
\begin{equation} G^* = \inf_{\alpha \in
\mathcal{\tilde{A}}\bigcap\mathbb{R}^{(n+k)+} }\Sigma_{i=1}^{n+k}
\alpha_i. \label{eqn:G*}
\end{equation}
We have $(a)$ because $(\log \snr)^{n+k}$ does not change the
diversity gain, $\exp\left(-\snr^{-\alpha_i}\right)$ decays
exponentially with \snr\ if $\alpha_i<0$,
$\exp\left(-\snr^{-\alpha_i}\right)$ is $e$ for $\alpha_i =0$ and
$\exp\left(-\snr^{-\alpha_i}\right)$ approaches 1 for $\alpha_i
>0$ at high \snr\ \cite{ZhengT03}, $(b)$ follows because at high
\snr\ $\mathcal{A}$ converges to $\mathcal{\tilde{A}}$, finally
$(c)$ is due to Laplace's method~\cite{ZhengT03}.

As a result $d_{\mathcal{C}_S'}(r,t) = G^*$. To solve the
optimization problem of (\ref{eqn:G*}). we first solve the
subproblems
\begin{equation*}
s_i \triangleq \inf_{\alpha \in
\mathcal{\tilde{A}}\bigcap\mathbb{R}^{(n+k)+} \bigcap
\mathcal{S}_i}\Sigma_{i=1}^{n+k} \alpha_i,
\end{equation*}
where
\begin{eqnarray*}
\mathcal{S}_1 &=& \{ (\alpha_1, \alpha_{n+1})|0 \leq \alpha_{n+1}
\leq 1 \leq \alpha_1\}\\
\mathcal{S}_2 & = & \{ (\alpha_1, \alpha_{n+1})|0 \leq \alpha_{1}
\leq
\alpha_{n+1} \leq 1\} \\
\mathcal{S}_3 & = & \{ (\alpha_1, \alpha_{n+1})|0 \leq \alpha_{n+1}
\leq
\alpha_{1} \leq 1\} \\
\mathcal{S}_4 & = & \{ (\alpha_1, \alpha_{n+1})|0 \leq
\alpha_{1}\leq 1
\leq \alpha_{n+1} \} \\
\mathcal{S}_5 & = & \{ (\alpha_1, \alpha_{n+1})|1 \leq \alpha_{1}
,\alpha_{n+1} \}.
\end{eqnarray*}

As an example, suppose we want to find $s_1$. Thus we have the
following linear optimization problem
\begin{eqnarray*}
\mbox{minimize~} \Sigma_{i=1}^{n+k} \alpha_i\\
t( 1-\alpha_{n+1})-r &\leq & 0 \\
0 ~\leq~ \alpha_{n+1} ~\leq ~1 &\leq& \alpha_1 \\
 \min \{\alpha_1, ...\alpha_n\} &=&\alpha_1 \\
 \min\{\alpha_{n+1}, ..., \alpha_{n+k}\} & = & \alpha_{n+1}
\end{eqnarray*}
This problem has two solutions at
\begin{eqnarray*}\lefteqn{(\alpha_1^*,...,\alpha_n^*, \alpha_{n+1}^*,
...,\alpha_{n+k}^*) }\\
&=& \left\{\begin{array}{lll}
  (1,...,1,0,...0) & \mbox{if} & t\leq r \\
  (1,...,1,1-r/t,...,1-r/t) & \mbox{if} & t \geq r \\
\end{array} \right..
\end{eqnarray*}
Then for $ \alpha \in
\mathcal{\tilde{A}}\bigcap\mathbb{R}^{(n+k)+}\bigcap \mathcal{S}_1$
\begin{equation*}s_1 = \min \Sigma \alpha_i =
\left\{\begin{array}{lll}
  n & \mbox{if} & t\leq r \\
  n+k(1-r/t) & \mbox{if} & t \geq r \\
\end{array} \right. .\label{eqn:s1}
\end{equation*}
Similarly, we find $s_2,s_3,s_4$ and $s_5$. Then $ G^* = \min_i
s_i$, which concludes the proof.

\section{Proof of Theorem~\ref{thm:MARC_CF_DMT}}
\label{app:MARC_CF_DMT}

When $S_1$ and $S_2$ do equal time sharing and $t = 1/2$, we use
Corollary~\ref{cor:HD_CF_111} to conclude that $d_{MARC,
CF}^{TS}(\bar{r}) = 2(1-r)$ is achievable, where $TS$ denotes time
sharing. Next, we discuss the case when both sources transmit
together.

In the half-duplex \marc, when both sources transmit simultaneously
and the relay does \cf\ for the signal it receives, similar to \cf\
discussed in Sections~ \ref{sec:MultiAntennaSingleRelay} and
\ref{sec:MultiAntennaSingleRelayHD}, the information rates satisfy
\begin{eqnarray*}
R^{({S_1})} &\leq& t I(X_{S_1};\hat{Y}_RY_ D|X_{S_2},q_1)\\
&&{ +}\:
(1-t)I(X_{S_1};Y_D|X_{S_2}X_R,q_2) \\
R^{({S_2})} &\leq& t I(X_{S_2};\hat{Y}_RY_ D|X_{S_1},q_1) \\
&&{ +}\:(1-t)I(X_{S_2};Y_D|X_{S_1}X_R,q_2) \\
R^{({S_1})}+R^{({S_2})} &\leq& t
I(X_{S_1}X_{S_2};\hat{Y}_RY_D|q_1)\\
&&{ +}\: (1-t) I(X_{S_1}X_{S_2};Y_D|X_R,q_2)
\end{eqnarray*}
for independent $X_{S_1}$, $X_{S_2}$, and $X_R$ subject to
\begin{eqnarray}
t I(\hat{Y}_R;Y_R|Y_D,q_1) \leq (1-t) I(X_R;Y_D|q_2),
\label{eqn:MARCCompressionConstraint}
\end{eqnarray}
where $\hat{Y}_R$ is the auxiliary random variable which denotes the
quantized signal at the relay and depends on $Y_R$ and
$X_R$~\cite{SankaranarayananKM04} and $t$ is the fraction of time
the relay listens.

To compute these mutual information, we assume $X_{S_1}$ and
$X_{S_2}$ are independent, complex Gaussian with zero mean, have
variances $P_{S_1}$ and $P_{S_2}$ respectively, and
$\hat{Y}_R=Y_{R,1}+\hat{Z}_R$, where $\hat{Z}_R$ is a complex
Gaussian random variable with zero mean and variance $\hat{N}_R$ and
is independent from all other random variables. We define
\begin{eqnarray*}
L_{S_1,D}  & \triangleq &
 1+|h_{S_1D}|^2P_{S_1}   \label{eqn:L_S1,D} \\
L_{S_2,D}  & \triangleq &   1+ |h_{S_2D}|^2P_{S_2}  \label{eqn:L_S2,D}\\
L_{S_1S_2,D}  & \triangleq & 1+ |h_{S_2D}|^2P_{S_1}+|h_{S_2D}|^2P_{S_2}  \label{eqn:L_S1S2,D}\\
L_{S_1S_2R,D} &  \triangleq &
1+|h_{S_1D}|^2P_{S_1}+|h_{S_2D}|^2P_{S_2}+|h_{RD}|^2P_{R}
\label{eqn:L_S1S2R,D} \\
L_{S_1,RD} & \triangleq & 1+|h_{S_1R}|^2P_{S_1}+|h_{S_1D}|^2P_{S_1} \nonumber \\&&{ +}\:\hat{N}_R(1+|h_{S_1D}|^2P_{S_1})\label{eqn:L_S1,RD} \\
L_{S_2,RD}  & \triangleq &  1+|h_{S_2R}|^2P_{S_2}+|h_{S_2D}|^2P_{S_2}\nonumber \\ &&{ +}\:\hat{N}_R(1+|h_{S_2D}|^2P_{S_2})\label{eqn:L_S2,RD} \\
L_{S_1S_2,RD} &  \triangleq &  \left|
 \bH_{S_1S_2,RD}\left[ \begin{array}{cc}
   P_{S_1} & 0 \\
   0 & P_{S_2} \\
 \end{array}\right]\bH_{S_1S_2,RD}^{\dag}\right.\\
 &&{ +}\:\left.\left[\begin{array}{cc} \hat{N}_R+1 & 0
\\ 0 & 1
\end{array} \right]
\right|\label{eqn:L_S1S2,RD}
\end{eqnarray*} where
$\bH_{S_1S_2,RD} =
\left[\begin{array}{cc}h_{S_1R}& h_{S_2R} \\
h_{S_1D} & h_{S_2D}\end{array}\right]$.

Since the relay has relevant \csi, using
(\ref{eqn:MARCCompressionConstraint}) it can choose the compression
noise variance $\hat{N}_R$ to satisfy
\begin{equation*}
\hat{N}_R=\frac{L_{S_1S_2,RD}}{U}, \mbox{~with~}  U = L_{S_1S_2,D}
\left(\frac{L_{S_1S_2R,D}}{L_{S_1S_2,D}}\right)^{(\frac{1-t}{t})}.
\end{equation*}
Then
\begin{eqnarray}
R^{({S_1})} &\leq& t \log \frac{L_{S_1,RD}}{\hat{N}_R+1}+
(1-t)\log L_{S_1,D} \label{eqn:rate1}\\
R^{({S_2})} &\leq& t \log \frac{L_{S_2,RD}}{\hat{N}_R+1}+
(1-t)\log L_{S_2,D} \nonumber\\
R^{({S_1})}+R^{({S_2})} &\leq& t \log
\frac{L_{S_1S_2,RD}}{\hat{N}_R+1}\nonumber \\
&&{ +}\: (1-t) \log L_{S_1S_2,D} \nonumber
\end{eqnarray}

To find a lower bound on the achievable \dmt, we use the union bound
on the probability of outage. For symmetric users with individual
target data rates $R^{(T_{S_1})}=R^{(T_{S_2})} = r/2 \log \snr$, and
a target sum data rate $R^{(T)}=r \log \snr$ the probability of
outage at the destination is
\begin{eqnarray}
\lefteqn{P(\mbox{outage at D})} \nonumber\\
 &\leq& P(R^{({S_1})} <
R^{(T)}/2)+P(R^{({S_2})}<R^{(T)}/2)
\nonumber \\
&&{ +}\:P(R^{({S_1})}+R^{({S_2})}<R^{(T)}).  \label{eqn:MARC_Pout}
\end{eqnarray}

One can prove that the first and second terms $P(R^{({S_1})} <
R^{(T)}/2)$ and $P(R^{({S_2})}<R^{(T)}/2)$ are on the order of
$\snr^{-(1-r/2)}$ at high \snr, for any $t$. To see this we write
(\ref{eqn:rate1}) explicitly as
\begin{eqnarray*}
R^{(S_1)} &\leq & t \log \left( 1 + |h_{S_1D}|^2P_{S_1} +
\frac{|h_{S_1R}|^2P_{S_1}}{\hat{N}_R+1}\right) \\&&{ +}\: (1-t) \log
\left( 1 + |h_{S_1D}|^2P_{S_1} \right)
\end{eqnarray*}
As the relay compresses both sources together, the compression noise
is on the order of \snr\ and the term
$|h_{S_1R}|^2P_{S_1}/(\hat{N}_R+1)$ does not contribute to the
overall mutual information at high \snr.

The last term in (\ref{eqn:MARC_Pout}) can be analyzed similar to
Section~\ref{sec:MultiAntennaSingleRelay}, as this term mimics the 2
antenna source, 1 antenna relay and 1 antenna destination behavior.
For $m_1 = m_2 = k=1$, we follow the proof from (\ref{eqn:CFproof}).
\begin{eqnarray*}
\lefteqn{P(R^{({S_1})}+R^{({S_2})}<R^{(T)})}\\
 &\leq & P \left(
{L_{S_1S_2,RD}^{\prime}}^tL_{S_1S_2,D}^{(1-t)} < 2^k\snr^{r}
\right)\\ &&{ +}\:P \left( L_{S_1S_2R,D}^{(1-t)}L_{S_1S_2,D}^{t} <
2^k\snr^{r} \right) \\
&\leq & P \left( L_{S_1S_2R,D}^tL_{S_1S_2,D}^{(1-t)} < 2^k\snr^{r}
\right)\\&&{ +}\: P \left( L_{S_1S_2R,D}^{(1-t)}L_{S_1S_2,D}^{t} <
2^k\snr^{r}
\right)  \\
& \dot{=} & \snr^{-d_{\mathcal{C}_D}'(r,t)} +\snr^{-d_{\mathcal{C}_D}'(r,1-t)}\\
& \dot{=} & \snr^{-\min \{d_{\mathcal{C}_D}'(r,t),
d_{\mathcal{C}_D}'(r,1-t)\}},
\end{eqnarray*}
From first line to the second, we used the fact that $
L_{S_1S_2,RD}^{\prime} \geq L_{S_1S_2R,D}$ with \[ L_{S_1S_2,RD}'
 \triangleq  \left|
 \bH_{S_1S_2,RD}\left[ \begin{array}{cc}
   P_{S_1} & 0 \\
   0 & P_{S_2} \\
 \end{array}\right]\bH_{S_1S_2,RD}^{\dag}+\mathbf{I}_2
\right|,\] as a $2 \times 2$ multiple antenna system has higher
capacity than a $3 \times 1$ system.

Using $d_{\mathcal{C}_D}'(r,t)$ from
Theorem~\ref{thm:HalfDuplexDMTBound} with $m= m_1+m_2 = 2$, $k=1$,
$n=1$, to maximize $\min \{d_{\mathcal{C}_D}'(r,t),
d_{\mathcal{C}_D}'(r,1-t)\}$ over $t$, we need to choose
$\frac{1}{3} \leq t \leq \frac{2}{3}$ and thus
\begin{eqnarray*}
d_{MARC, CF}^{SIM}(\bar{r}) & \geq & \min \left \{ 1-\frac{r}{2},
3(1-r) \right \},
\end{eqnarray*} where $SIM$ denotes simultaneous transmisssion.

To find an upper bound on the achievable \dmt\ we write
\begin{eqnarray*}\lefteqn{P(\mbox{outage at D}) }\\&\geq& \max\{P(R^{({S_1})} <
R^{(T)}/2),P(R^{({S_2})}<R^{(T)}/2)\}, \end{eqnarray*} so \[
d_{MARC,CF}^{SIM}(\bar{r}) \leq 1-\frac{r}{2}.\] Combining this with
the upper bound in (\ref{eqn:marc_upperbound}), and with
$d_{MARC,CF}^{TS}(\bar{r})$, we have Theorem~\ref{thm:MARC_CF_DMT}.

\section{Proof of Upper Bound in Theorems~\ref{thm:SingleAntennaTwoRelayNonClus} and
\ref{thm:SingleAntennaTwoRelayClus}}\label{app:SingleAntennaTwoRelayUB}

To provide upper bounds, we will use the cut-set bounds as argued in
Lemma~\ref{Lemma:CutsetUpperBound}. The cut-sets of interest are
shown in Fig.~\ref{fig:SingleAntennaTwoRelay} and denoted as
$\mathcal{C}_S$, $\mathcal{C}_{SR_1}$ and $\mathcal{C}_D$. We will
see that these will be adequate to provide a tight bound.

In order to calculate the diversity orders $d_{\mathcal{C}_i}(r)$
for each cut-set, we write down the instantaneous mutual information
expressions given the fading levels as
\begin{eqnarray*}
I_{\mathcal{C}_S}& =& I(X_S;Y_{R_1}Y_{R_2}Y_D|X_{R_1}X_{R_2}) \label{eqn:SingleAntennaTwoRelayCutset1}\\
I_{\mathcal{C}_{SR_1}} & =& I(X_SX_{R_1};Y_{R_2}Y_D|X_{R_2}) \label{eqn:SingleAntennaTwoRelayCutset2}\\
I_{\mathcal{C}_D} & =&I(X_SX_{R_1}X_{R_2};Y_D).
\label{eqn:SingleAntennaTwoRelayCutset4}
\end{eqnarray*}

To maximize this upper bound we need to choose $X_S$, $X_{R_1}$ and
$X_{R_2}$ complex Gaussian with zero mean and variances $P_S$,
$P_{R_1}$ and $P_{R_2}$ respectively, where $P_S$, $P_{R_1}$ and
$P_{R_2}$ denote the average power constraints each node
has~\cite{TseV05}. Then
\begin{eqnarray}
I_{\mathcal{C}_{S}} &\leq & I_{\mathcal{C}_{S}}^\prime \nonumber\\
& =& \log \left( 1+
|a_{SR_1}|^2P_{S}+|h_{SR_2}|^2P_{S}+|h_{SD}|^2P_{S}
\right) \nonumber\\ \label{eqn:SingleAntennaTwoRelayNonClusteredCutset1}\\
I_{\mathcal{C}_{SR_1}} &\leq & I_{\mathcal{C}_{SR_1}}^\prime \nonumber\\
& =&  \log \left|  \mathbf{I}_2+
\bH_{SR_1,R_2D}\bH_{SR_1,R_2D}^\dagger
(P_S+P_{R_1}) \right| \nonumber\\   \label{eqn:SingleAntennaTwoRelayNonClusteredCutset2}\\
I_{\mathcal{C}_{D}} &\leq & I_{\mathcal{C}_{D}}^\prime \nonumber\\
& =&  \log \left( 1+ \left(|h_{SD}|^2+|h_{R_1D}|^2+|a_{R_2D}|^2
\right) \right. \nonumber \\&&{.}\: \left.
\left(P_{S}+P_{R_1}+P_{R_2}\right)\right)
\label{eqn:SingleAntennaTwoRelayNonClusteredCutset4}
\end{eqnarray}
where
\begin{equation}
{\bH}_{SR_1,R_2D} = \left[%
\begin{array}{cc}
  h_{SR_2} &   h_{R_1R_2} \\
  h_{SD} &   h_{R_1D} \\
\end{array}%
\right] \label{eqn:matrix_A}
\end{equation} and we used
(\ref{eqn:covarianceremoval}) to upper bound
$I_{\mathcal{C}_{SR_1}}$ with $I_{\mathcal{C}_{SR_1}}^\prime$ in
(\ref{eqn:SingleAntennaTwoRelayNonClusteredCutset2}) and
$I_{\mathcal{C}_{D}}$ with $I_{\mathcal{C}_{D}}^\prime$ in
(\ref{eqn:SingleAntennaTwoRelayNonClusteredCutset4}).

For a target data rate $R^{(T)} = r \log \snr$,
$P(I_{\mathcal{C}_{S}}^\prime < R^{(T)}) \dot{=}
P(I_{\mathcal{C}_D}^\prime <R^{(T)}) \dot{=} \snr^{-d_{13}(r)}$,
whereas $P(I_{\mathcal{C}_{SR_1}}^\prime< R^{(T)}) \dot{=}
\snr^{-d_{22}(r)}$. Then using Lemma~\ref{Lemma:CutsetUpperBound},
the best achievable diversity $d(r)$ of a non-clustered system is
upper bounded by $d(r) \leq \min \{ d_{13}(r), d_{22}(r)\} =
d_{13}(r)$.

When the system is clustered, $I_{\mathcal{C}_{S}}^\prime$ and
$I_{\mathcal{C}_{D}}^\prime$ are larger than the Gaussian channel
capacities $ \log \left( 1+ G_{SR_1}P_{S} \right)$ and $\log \left(
1+ G_{R_2D}P_{R_2} \right)$ respectively. Then
$P(I_{\mathcal{C}_{S}}^\prime < R^{(T)} ) \dot{=}
P(I_{\mathcal{C}_{D}}^\prime < R^{(T)}) \dot{=} \snr^{-\infty}$, if
$r \leq 1$. In other words, it is possible to operate at the
positive rate of $R^{(T)}$ reliably without any outage and as $\snr$
increases, the data rate of this bound can increase as $\log \snr$
without any penalty in reliability. However, this is not the case
for any $r> 1$ as $P(I_{\mathcal{C}_{S}}^\prime= < R^{(T)} ) \dot{=}
P(I_{\mathcal{C}_{D}}^\prime < R^{(T)}) \dot{=} 1$. Combining these
results with the upper bound due to $I_{\mathcal{C}_{SR_1}}$, we
have
\[ d(r) \leq \left \{\begin{array}{lll}
  d_{22}(r) & \mbox{if} & r \leq 1 \\
  0 & \mbox{if} & r > 1 \\
\end{array}\right. .\]

\section{Proof of Achievability in
Theorem~\ref{thm:SingleAntennaTwoRelayNonClus}}\label{app:SingleAntennaTwoRelayNonClus}

We assume the source, $R_1$ and $R_2$ perform block Markov
superposition coding. After each block $R_1$ and $R_2$ attempt to
decode the source. The destination does backward decoding similar to
the case in Section~\ref{sec:MultiAntennaSingleRelay}.

Using the block Markov coding structure both $R_1$ and $R_2$ can
remove each other's signal from its own received signal in
(\ref{eqn:SingAntTwoRelayR1receivedsignal}) and
(\ref{eqn:SingAntTwoRelayR2receivedsignal}) before trying to decode
any information.

We choose ${X_{R_1}}$ and ${X_{R_2}}$ independent complex Gaussian
with zero mean and variances $P_{R_1}$ and $P_{R_2}$ respectively.
We also choose ${X_{S}}$ independently with complex Gaussian
distribution $\mathcal{CN}(0,P_{S})$. Then the probability of outage
for this system, when the target data rate $R^{(T)} = r \log \snr$,
is equal to
\begin{eqnarray}
\lefteqn{P(\mbox{outage at D})} \nonumber \\
&=& P(\mbox{outage and both relays decode})\nonumber \\
&&{ +}\:
P(\mbox{outage and~}R_1 \mbox{~decodes,~} R_2 \mbox{~cannot decode}) \nonumber\\
&&{+}\:P(\mbox{outage and~}R_1 \mbox{~cannot decode, and~} R_2
\mbox{~decodes})\nonumber \\
&&{+}\: P(\mbox{outage, and none of the relays decode})
\label{eqn:SingleAntennaTwoRelayNonClusteredDF}.
\end{eqnarray}
\begin{eqnarray*}
\lefteqn{P(\mbox{outage at D})} \nonumber \\& = &P(L_{SR_1R_2,D}<
\snr^r , L_{S,R_1}> \snr^r ,L_{S,R_2}>\snr^r)  \nonumber\\
&&{+}\:P(L_{SR_1,D}<
\snr^r , L_{S,R_1}>\snr^r ,L_{S,R_2}<\snr^r) \nonumber \\
&&{+}\:P(L_{SR_2,D}<
\snr^r , L_{S,R_1}<\snr^r ,L_{S,R_2}>\snr^r) \nonumber \\
&&{+}\:P(L_{S,D}< \snr^r , L_{S,R_1}< \snr^r ,L_{S,R_2}<
\snr^r),\nonumber
\end{eqnarray*}
where
\begin{eqnarray}
L_{S,R_1} & \triangleq & \log\left( 1+|a_{SR_1}|^2P_{S}\right) \label{eqn:L_S,R1}\\
L_{S,R_2} &\triangleq&\log\left( 1+|h_{SR_2}|^2P_{S}\right) \nonumber\\
L_{SR_1R_2,D}&\triangleq&  \log \left(1+|h_{SD}|^2P_{S}+|h_{R_1D}|^2P_{R_1}\right.\nonumber\\
&&{+}\: \left.|a_{R_2D}|^2P_{R_2}\right)\label{eqn:L_SR1R2,D}\\
L_{SR_1,D}&\triangleq&  \log \left( 1+|h_{SD}|^2P_{S}+|h_{R_1D}|^2P_{R_1}\right)\label{eqn:L_SR1,D}\\
L_{SR_2,D}&\triangleq&  \log \left(1+|h_{SD}|^2P_{S}+|h_{R_2D}|^2P_{R_2}\right)\nonumber\\
L_{S,D}&\triangleq&  \log \left(1+ |h_{SD}|^2P_{S} \right),\nonumber
\end{eqnarray}
and $a_{SR_1} = h_{S,R_1}$, $a_{R_2D} = h_{R_2D}$ as the system is
non-clustered.

Using the fact that $P(L_{S,R_1}> \snr^r)\dot{=}1$ and
$P(L_{S,R_2}>\snr^r)\dot{=}1$, this outage probability becomes
\begin{eqnarray*} \lefteqn{P(\mbox{outage at D})}\\ &\dot{=}&
\snr^{-d_{31}(r)}+2 \snr^{-d_{21}(r)}\snr^{-d_{11}(r)}\\
&&{+}\: \snr^{-d_{11}(r)}\snr^{-d_{11}(r)}\snr^{-d_{11}(r)}\\
&\dot{=}&\snr^{-d_{31}(r)}
\end{eqnarray*}at high \snr, which is equivalent to the outage behavior of a $1 \times 3$
system (or $3 \times 1$) system. Hence, in a non-clustered system if
both relays do \df, the \dmt\ in
Theorem~\ref{thm:SingleAntennaTwoRelayNonClus} can be achieved.

\section{Proof of Achievability in
Theorem~\ref{thm:SingleAntennaTwoRelayClus}}\label{app:SingleAntennaTwoRelayClus}
To prove that the \dmt\ of
Theorem~\ref{thm:SingleAntennaTwoRelayClus} is achievable, we use
the mixed strategy suggested in~\cite{KramerGG04}, in which $R_1$
does \df\ and then the source node and the first relay together
perform block Markov superposition encoding. Similar to the
non-clustered case in
Appendix~\ref{app:SingleAntennaTwoRelayNonClus}, we require $R_1$ to
decode the source message reliably, and to transmit only if this is
the case. We assume that $R_2$ and the destination know if $R_1$
transmits or not. The second relay $R_2$ does \cf.

To prove the \dmt\ we calculate the probability of outage as
\begin{eqnarray*}
P(\mbox{outage at D}) &=& P(\mbox{outage}|R_1 \mbox{~decodes})P(R_1
\mbox{~decodes}) \nonumber
\\ &&{+}\: P(\mbox{outage}|R_1 \mbox{~cannot decode})\\
&&{.}\: P(R_1 \mbox{~cannot decode}).
\end{eqnarray*}

As $R_1$ is clustered with the source, source to $R_1$ communication
is reliable for all multiplexing gains up to 1; i.e. $P(R_1
\mbox{~decodes})$ and $P(R_1 \mbox{~cannot decode})$ can be made
arbitrarily close to 1 and 0 respectively. Therefore, we only need
to show that $P(\mbox{outage}|R_1 \mbox{~decodes})$ decays at least
as fast as $d_{22}(r)$ with increasing $\snr$.

When $R_1$ decodes the source message reliably, if the target data
rate $R^{(T)}$ satisfies
\begin{eqnarray}
R^{(T)} <
I(X_{S}X_{R_1};\hat{Y}_{R_2}Y_{D}|X_{R_2})\label{eqn:CFrate_with_R1}
\end{eqnarray}
subject to
\begin{eqnarray}
I(\hat{Y}_{R_2};Y_{R_2}|X_{R_2}Y_D)\leq I(X_{R_2};Y_D),
\label{eqn:SingleAntennaTwoRelayCompressionConstraint}
\end{eqnarray}
then the system is not in outage.

We choose $X_S$, $X_{R_1}$ and $X_{R_2}$ independent complex
Gaussian with variances $P_S$, $P_{R_1}$ and $P_{R_2}$ respectively
and $\hat{Y}_{R_2}=Y_{R_2}+\hat{Z}_{R_2}$, where $\hat{Z}_{R_2}$ is
an independent complex Gaussian random variable with zero mean,
variance $\hat{N}_{R_2}$ and independent from all other random
variables.

We define
\begin{eqnarray*}
L_{SR_1,R_2D} & \triangleq &
\left|\bH_{SR_1,R_2D}\left[\begin{array}{cc}
  P_S & 0 \\
  0 & P_{R_1} \\
\end{array}\right]\bH_{SR_1,R_2D}^{\dag}\right. \nonumber \\&&{+}\: \left. \left[ \begin{array}{cc}
\hat{N}_{R_2}+1 & 0 \\ 0 & 1
\end{array}\right]\right|, \label{eqn:L_SR1,R2D} \\
L_{SR_1,R_2D}'  & \triangleq & \left|\bH_{SR_1,R_2D}
\left[\begin{array}{cc}
  P_S & 0 \\
  0 & P_{R_1} \\
\end{array}\right]\bH_{SR_1,R_2D}^{\dag}+\mathbf{I}_2\right|, \nonumber
\\ \label{eqn:L_SR1,R2D'}
\end{eqnarray*} where ${\bH_{SR_1,R_2D}}$ is given in
(\ref{eqn:matrix_A}). Using the definitions of $L_{S,R_1}$,
$L_{SR_1,D}$, and $L_{SR_1R_2,D}$ from (\ref{eqn:L_S,R1}),
(\ref{eqn:L_SR1,D}) and (\ref{eqn:L_SR1R2,D}), with $a_{S,R_1} = G$
and $a_{R_2,D} = G$, the instantaneous mutual information
expressions conditioned on the fading levels for the mixed strategy
become
\begin{eqnarray*}
I(X_S;Y_{R_1}|X_{R_1})&=&\log L_{S,R_1}\\
I(X_SX_{R_1};\hat{Y}_{R_2}Y_D|X_{R_2}) & = & \log
\frac{L_{SR_1,R_2D}}{{1+\hat{N}_{R_2}}}.\label{eqn:SingleAntennaTwoRelayMixedRateCalculated}
\end{eqnarray*}
The mutual information in the compression rate constraint of
(\ref{eqn:SingleAntennaTwoRelayCompressionConstraint}) are
\begin{eqnarray*}
 I(\hat{Y}_{R_2};Y_{R_2}|X_{R_2}Y_D)
& =&\log \frac{L_{SR_1,R_2D}}{L_{SR_1,D}\hat{N}_{R_2}} \label{eqn:CompressionRate}\\
 I(X_{R_2};Y_D) &=& \log \frac{L_{SR_1R_2,D}}{L_{SR_1,D}} \label{eqn:R2_DchannelRate}.
\end{eqnarray*}
Then the compression noise power has to be chosen to satisfy
\begin{eqnarray}
\hat{N}_{R_2} &\geq&\frac{L_{SR_1,R_2D}}{L_{SR_1R_2,D}}.
\label{eqn:SingleAntennaTwoRelayClusteredHatN3}
\end{eqnarray}
Note that both sides of the above inequality are functions of
$\hat{N}_{R_2}$. Using the \csi, the relay will always ensure
(\ref{eqn:SingleAntennaTwoRelayClusteredHatN3}) is satisfied.

After substituting the value of the compression noise in
(\ref{eqn:CFrate_with_R1}) we need to calculate $P(\mbox{outage}|R_1
\mbox{~decodes})$. Given $R_1$ decodes, this problem becomes similar
to \textit{Problem 1}, and we can find that when
$P(\mbox{outage}|R_1 \mbox{~decodes}) \dot{=} \snr^{-d(r)}$, $d(r) =
d_{22}(r)$. Finally, as $P(\mbox{outage at D}) \dot{=}
\snr^{-d(r)}$, we say the mixed strategy achieves the \dmt\ bound.

\section*{Acknowledgment}
The authors would like to thank Dr.~Gerhard Kramer, whose comments
improved the results, and the guest editor Dr.~J.~Nicholas Laneman
and the anonymous reviewers for their help in the organization of
the paper.


\begin{thebibliography}{10}
\providecommand{\url}[1]{#1} \csname url@samestyle\endcsname
\providecommand{\newblock}{\relax} \providecommand{\bibinfo}[2]{#2}
\providecommand{\BIBentrySTDinterwordspacing}{\spaceskip=0pt\relax}
\providecommand{\BIBentryALTinterwordstretchfactor}{4}
\providecommand{\BIBentryALTinterwordspacing}{\spaceskip=\fontdimen2\font
plus \BIBentryALTinterwordstretchfactor\fontdimen3\font minus
  \fontdimen4\font\relax}
\providecommand{\BIBforeignlanguage}[2]{{%
\expandafter\ifx\csname l@#1\endcsname\relax
\typeout{** WARNING: IEEEtran.bst: No hyphenation pattern has been}%
\typeout{** loaded for the language `#1'. Using the pattern for}%
\typeout{** the default language instead.}%
\else \language=\csname l@#1\endcsname \fi #2}}
\providecommand{\BIBdecl}{\relax} \BIBdecl

\bibitem{Rappaport}
T.~S. Rappaport, \emph{{W}ireless {C}ommunications: {P}rinciples
{\&}
  {P}ractice}, {S}econd~ed.\hskip 1em plus 0.5em minus 0.4em\relax Prentice
  Hall, 2002.

\bibitem{Foschini96}
G.~J. Foschini, ``Layered space-time architecture for wireless
communication in
  a fading environment when using multi-element antennas,'' \emph{Bell
  Laboratories Technical Journal}, p.~41, October 1996.

\bibitem{Telatar99}
I.~E. Telatar, ``Capacity of multiple-antenna {G}aussian channels,''
  \emph{Europian Transactions on Telecommunications}, vol.~10, p. 585, November
  1999.

\bibitem{SendonarisEA03_1}
A.~Sendonaris, E.~Erkip, and B.~Aazhang, ``User cooperation
diversity-{P}art
  {I}: {S}ystem description,'' \emph{IEEE Transactions on Communications},
  vol.~51, no.~11, p. 1927, November 2003.

\bibitem{SendonarisEA03_2}
------, ``User cooperation diversity-{P}art {II}: {I}mplementation aspects and
  performance analysis,'' \emph{IEEE Transactions on Communications}, vol.~51,
  no.~11, p. 1939, November 2003.

\bibitem{LanemanTW02}
J.~N. Laneman, D.~N.~C. Tse, and G.~W. Wornell, ``Cooperative
diversity in
  wireless networks: {E}fficient protocols and outage behavior,'' \emph{IEEE
  Transactions on Information Theory}, vol.~50, no.~12, p. 3062, December 2004.

\bibitem{VanDerMeulen71}
E.~C. van~der Meulen, ``Three-terminal communication channels,''
\emph{Advances
  in Applied Probability}, vol.~3, p. 121, 1971.

\bibitem{CoverElG79}
T.~M. Cover and A.~E. Gamal, ``Capacity theorems for the relay
channel,''
  \emph{IEEE Transactions on Information Theory}, vol.~25, no.~5, p. 572,
  September 1979.

\bibitem{BoyerFY04}
J.~Boyer, D.~D. Falconer, and H.~Yanikomeroglu, ``Multihop diversity
in
  wireless relaying channels,'' \emph{IEEE Transactions on Communications},
  vol.~52, no.~10, p. 1820, October 2004.

\bibitem{GastparV02}
M.~Gastpar and M.~Vetterli, ``On the capacity of wireless networks:
{T}he relay
  case,'' in \emph{Proceedings of IEEE INFOCOM}, 2002.

\bibitem{GuptaK03}
P.~Gupta and P.~R. Kumar, ``Towards an information theory of large
networks:
  {A}n achievable rate region,'' \emph{IEEE Transactions on Information
  Theory}, vol.~49, no.~2, p. 1877, August 2003.

\bibitem{HasnaA022}
M.~O. Hasna and M.-S. Alouini, ``Performance analysis of two-hop
relayed
  transmissions over {R}ayleigh fading channels,'' in \emph{IEEE Vehicular
  Technology Conference 2002}.

\bibitem{HostMadsen06}
A.~H{\o}st-Madsen, ``Capacity bounds for cooperative diversity,''
\emph{IEEE
  Transactions on Information Theory}, vol.~52, no.~4, p. 1522, April 2006.

\bibitem{HostMadsenZ05}
A.~H{\o}st-Madsen and J.~Zhang, ``Capacity bounds and power
allocation for
  wireless relay channels,'' \emph{IEEE Transactions on Information Theory},
  vol.~51, no.~6, p. 2020, June 2005.

\bibitem{JananiHHN04}
M.~Janani, A.~Hedayat, T.~E. Hunter, and A.~Nosratinia, ``Coded
cooperation in
  wireless communications: space-time transmission and iterative decoding,''
  \emph{IEEE Transactions on Signal Processing}, vol.~52, no.~2, p. 362,
  February 2004.

\bibitem{JindalMG04}
N.~Jindal, U.~Mitra, and A.~J. Goldsmith, ``Capacity of ad-hoc
networks with
  node cooperation,'' in \emph{Proceedings of IEEE International Symposium on
  Information Theory}, 2004, p. 271.

\bibitem{KatzShamai05}
M.~Katz and S.~Shamai, ``Transmitting to colocated users in wireless
ad hoc and
  sensor networks,'' \emph{IEEE Transactions on Information Theory}, vol.~51,
  no.~10, p. 3540, 2005.

\bibitem{KatzShamai06}
------, ``Relaying protocols for two co-located users,'' \emph{IEEE
  Transactions on Information Theory}, vol.~52, no.~6, p. 2329, 2006.

\bibitem{KhojastepourSA03}
M.~Khojastepour, A.~Sabharwal, and B.~Aazhang, ``On the capacity of
`cheap'
  relay networks,'' in \emph{Proceedings of 37th Conference on Information
  Sciences and Systems}, 2003.

\bibitem{KramerGG04}
G.~Kramer, M.~Gastpar, and P.~Gupta, ``Cooperative strategies and
capacity
  theorems for relay networks,'' \emph{IEEE Transactions on Information
  Theory}, vol.~51, no.~9, p. 3037, 2005.

\bibitem{LaiLiuEG05}
L.~Lai, K.~Liu, and H.~E. Gamal, ``On the achievable rate of
three-node
  wireless networks,'' in \emph{Proceedings of 2005 WirelessCom}, June 2005.

\bibitem{LaiLiuEG06}
------, ``The three node wireless network: Achievable rates and cooperation
  strategies,'' \emph{IEEE Transactions on Information Theory}, vol.~52, no.~3,
  p. 805, March 2006.

\bibitem{LanemanW03}
J.~N. Laneman and G.~W. Wornell, ``Distributed space-time coded
protocols for
  exploiting cooperative diversity in wireless networks,'' \emph{IEEE
  Transactions on Information Theory}, vol.~49, no.~10, p. 2415, October 2003.

\bibitem{LiangK06}
Y.~Liang and G.~Kramer, ``Rate regions for relay broadcast
channels,'' June
  2006.

\bibitem{LiangV05}
Y.~Liang and V.~V. Veeravalli, ``Cooperative relay broadcast
channels,'' March
  2007, iEEE Transactions on Information Theory, To appear.

\bibitem{MaricYK05}
I.~Maric, R.~D. Yates, and G.~Kramer, ``The discrete memoryless
compound
  multiple access channel with conferencing encoders,'' in \emph{Proceedings of
  IEEE International Symposium on Information Theory}, 2005.

\bibitem{NabarBolcskeiK04}
R.~U. Nabar, H.~Bolcskei, and F.~W. Kneubuhler, ``Fading relay
channels:
  {P}erformance limits and space-time signal design,'' \emph{IEEE Journal on
  Selected Areas in Communications}, vol.~22, no.~6, p. 1099, August 2004.

\bibitem{NgGoldsmith04}
C.~T.~K. Ng and A.~J. Goldsmith, ``Transmitter cooperation in ad-hoc
wireless
  networks: {D}oes dirty-paper coding beat relaying?'' in \emph{Proceedings of
  IEEE Information Theory Workshop}, 2004.

\bibitem{NgGoldsmith05}
------, ``Capacity gain from transmitter and receiver cooperation,'' in
  \emph{Proceedings of IEEE International Symposium on Information Theory},
  2005.

\bibitem{NgLanemanGoldsmith06}
C.~T.~K. Ng, J.~N. Laneman, and A.~J. Goldsmith, ``The role of {SNR}
in
  achieving {MIMO} rates in cooperative systems,'' in \emph{Proceedings of IEEE
  Information Theory Workshop}, 2006.

\bibitem{ReznikKV04}
A.~Reznik, S.~R. Kulkarni, and S.~Verdu, ``Degraded {G}aussian
multirelay
  channel: capacity and optimal power allocation,'' \emph{IEEE Transactions on
  Information Theory}, vol.~50, no.~12, p. 3037, December 2004.

\bibitem{ScheinG00}
B.~Schein and R.~Gallager, ``The {G}aussian parallel relay
network,'' in
  \emph{Proceedings of IEEE International Symposium on Information Theory},
  2000.

\bibitem{SheaWAL03_1}
J.~M. Shea, T.~F. Wong, A.~Avudainayagam, and X.~Li, ``Reliability
exchange
  schemes for iterative packet combining in distributed arrays,'' in \emph{IEEE
  Wireless Communications and Networking Conference}, 2003, p. 832.

\bibitem{YukselE03}
M.~Yuksel and E.~Erkip, ``Diversity in relaying protocols with
amplify and
  forward,'' in \emph{Proceedings of IEEE GLOBECOM}, 2003.

\bibitem{YukselE04_2}
------, ``Broadcast strategies for the fading relay channel,'' in
  \emph{Proceedings of IEEE MILCOM 2004}, 2004.

\bibitem{ZahediMElG04}
S.~Zahedi, M.~Mohseni, and A.~E. Gamal, ``On the capacity of {AWGN}
relay
  channels with linear relaying functions,'' in \emph{Proceedings of IEEE
  International Symposium on Information Theory}, 2004, p. 399.

\bibitem{ZhengT03}
L.~Zheng and D.~N.~C. Tse, ``Diversity and multiplexing: A
fundamental tradeoff
  in multiple-antenna channels,'' \emph{IEEE Transactions on Information
  Theory}, vol.~49, p. 1073, May 2003.

\bibitem{YukselE04}
M.~Yuksel and E.~Erkip, ``Diversity gains and clustering in wireless
  relaying,'' in \emph{Proceedings of IEEE International Symposium on
  Information Theory}, 2004, p. 400.

\bibitem{AzarianEGS05}
K.~Azarian, H.~El-Gamal, and P.~Schniter, ``On the achievable
  diversity-multiplexing tradeoff in half-duplex cooperative channels,''
  \emph{IEEE Transactions on Information Theory}, vol.~51, no.~12, p. 4152,
  December 2005.

\bibitem{YangBelfiore06}
S.~Yang and J.-C. Belfiore, ``Towards the optimal
amplify-and-forward
  cooperative diversity scheme,'' March 2006, iEEE Transactions on Information
  Theory, Accepted.

\bibitem{WangZHostMadsen05}
B.~Wang, J.~Zhang, and A.~H{\o}st-Madsen, ``On the capacity of
{MIMO} relay
  channels,'' \emph{IEEE Transactions on Information Theory}, vol.~51, p.~29,
  January 2005.

\bibitem{YangBelfiore07}
S.~Yang and J.-C. Belfiore, ``Optimal space-time codes for the
{MIMO}
  amplify-and-forward cooperative channel,'' \emph{IEEE Transactions on
  Information Theory}, vol.~53, p. 647, February 2007.

\bibitem{KramervWijngaarden00}
G.~Kramer and A.~J. van Wijngaarden, ``On the white {G}aussian
multiple-access
  relay channel,'' in \emph{Proceedings of IEEE International Symposium on
  Information Theory}, 2000.

\bibitem{SankaranarayananKM04}
L.~Sankaranarayanan, G.~Kramer, and N.~B. Mandayam, ``Hierarchical
sensor
  networks: capacity bounds and cooperative strategies using the
  multiple-access relay channel model,'' in \emph{Proceedings of First Annual
  IEEE Communications Society Conference on Sensor and Ad Hoc Communications
  and Networks}, 2004, p. 191.

\bibitem{AzarianEGS06}
K.~Azarian, H.~El-Gamal, and P.~Schniter, ``On the optimality of the
{ARQ-DDF}
  protocol,'' January 2006, IEEE Transactions on Information Theory, Accepted.

\bibitem{ChenL06}
D.~Chen and J.~N. Laneman, ``The diversity-multiplexing tradeoff for
the
  multiaccess relay channel,'' in \emph{Proceedings of 40th Conference of
  Information Sciences and Systems}, 2006.

\bibitem{ChenAL07}
D.~Chen, K.~Azarian, and J.~N. Laneman, ``A case for amplify-forward
relaying
  in the block-fading multiaccess channel,'' submitted to IEEE Transactions on
  Information Theory.

\bibitem{PrasadV04}
N.~Prasad and M.~K. Varanasi, ``Diversity and multiplexing tradeoff
bounds for
  cooperative diversity protocols,'' in \emph{Proceedings of IEEE International
  Symposium on Information Theory}, 2004, p. 268.

\bibitem{BletsasKRL06}
A.~Bletsas, A.~Khisti, D.~P. Reed, and A.~Lippman, ``A simple
cooperative
  diversity method based on network path selection,'' \emph{IEEE Journal on
  Selected Areas in Communications}, vol.~24, no.~3, p. 659, March 2006.

\bibitem{HostMadsenN05}
A.~H{\o}st-Madsen and A.~Nosratinia, ``The multiplexing gain of
wireless
  networks,'' in \emph{Proceedings of IEEE International Symposium on
  Information Theory}, 2005.

\bibitem{ZhaoMMW06}
L.~Zhao, W.~Mo, Y.~Ma, and Z.~Wang, ``Diversity and multiplexing
tradeoff in
  general fading channels,'' \emph{IEEE Transactions on Information Theory},
  vol.~53, p. 1549, April 2007.

\bibitem{ShinCL06}
S.~C. W.~Shin and Y.~H. Lee, ``Outage analysis for {MIMO} {R}ician
channels and
  channels with partial {CSI},'' in \emph{Proceedings of IEEE International
  Symposium on Information Theory}, 2006.

\bibitem{CaireTB99}
G.~Caire, G.~Taricco, and E.~Biglieri, ``Optimum power control over
fading
  channels,'' \emph{IEEE Transactions on Information Theory}, vol.~45, no.~5,
  p. 1468, July 1999.

\bibitem{KhoshnevisSabharwal04}
A.~Khoshnevis and A.~Sabharwal, ``On diversity and multiplexing gain
of
  multiple antennas systems with transmitter channel information,'' in
  \emph{Proceedings of 42nd Allerton Conference on Communication, Control and
  Computing}, 2004.

\bibitem{Cover}
T.~M. Cover and J.~A. Thomas, \emph{{E}lements of {I}nformation
  {T}heory}.\hskip 1em plus 0.5em minus 0.4em\relax John-Wiley {\&} Sons, Inc.,
  1991.

\bibitem{HornJ85}
R.~A. Horn and C.~R. Johnson, \emph{Matrix Analysis}.\hskip 1em plus
0.5em
  minus 0.4em\relax Cambridge University Press, 1985.

\bibitem{TseV05}
D.~Tse and P.~Viswanath, \emph{{F}undamentals of {W}ireless
  {C}ommunication}.\hskip 1em plus 0.5em minus 0.4em\relax Cambridge University
  Press, 2005.

\bibitem{ZengKBuzo89}
C.~Zeng, F.~Kuhlmann, and A.~Buzo, ``Achievability proof of some
multiuser
  channel coding theorems using backward decoding,'' \emph{IEEE Transactions on
  Information Theory}, vol.~35, no.~6, p. 1160, November 1989.

\bibitem{WillemsvM85}
F.~M.~J. Willems and E.~C. van~der Meulen, ``The discrete memoryless
  multiple-access channel with cribbing encoders,'' \emph{IEEE Transactions on
  Information Theory}, vol.~31, p. 313, May 1985.

\bibitem{Kramer04}
G.~Kramer, ``Models and theory for relay channels with receive
constraints,''
  in \emph{Proceedings of 42nd Allerton Conference on Communication, Control
  and Computing}, 2004.

\bibitem{TseVZ04}
D.~N.~C. Tse, P.~Viswanath, and L.~Zheng, ``Diversity-multiplexing
tradeoff in
  multiple-access channels,'' \emph{IEEE Transactions on Information Theory},
  vol.~50, p. 1859, September 2004.

\bibitem{JafarG05}
S.~A. Jafar and A.~J. Goldsmith, ``Isotropic fading vector broadcast
channels:
  The scalar upper bound and loss in degrees of freedom,'' \emph{IEEE
  Transactions on Information Theory}, vol.~51, no.~3, p. 848, 2005.

\bibitem{XieKumar04}
L.~Xie and P.~R. Kumar, ``A network information theory for wireless
  communication: Scaling laws and optimal operation,'' \emph{IEEE Transactions
  on Information Theory}, vol.~50, no.~5, p. 748, May 2004.

\bibitem{YukselE06_2}
M.~Yuksel and E.~Erkip, ``Diversity-multiplexing tradeoff in
multiple-antenna
  relay systems,'' in \emph{Proceedings of IEEE International Symposium on
  Information Theory}, 2006.

\bibitem{YukselE06}
------, ``Diversity-multiplexing tradeoff in cooperative wireless systems,'' in
  \emph{Proceedings of 40th Conference of Information Sciences and Systems},
  2006.

\bibitem{YukselE06_3}
------, ``Diversity-multiplexing tradeoff in half-duplex relay systems,'' in
  \emph{Proceedings of IEEE International Conference on Communications}, 2007.

\end{thebibliography}

\begin{biographynophoto}{Melda Yuksel}
[S'98] received her B.S. degree in Electrical and Electronics
Engineering from Middle East Technical University, Ankara, Turkey,
in 2001. She is currently working towards her Ph.D. degree at
Polytechnic University, Brooklyn, NY. In 2004, she was a summer
researcher in Mathematical Sciences Research Center, Bell-Labs,
Lucent Technologies, Murray Hill, NJ.

Melda Yuksel is the recipient of the best paper award in the
Communication Theory Symposium of ICC 2007. Her research interests
include communication theory and information theory and more
specifically cooperative communications, network information theory
and information theoretic security over communication channels.
\end{biographynophoto}

\begin{biographynophoto}{Elza Erkip}
[S'93, M'96, SM'05] received the Ph.D. and M.S. degrees in
Electrical Engineering from Stanford University, and the B.S. degree
in Electrical and Electronics Engineering from Middle East Technical
University, Turkey. She joined Polytechnic University in Spring
2000, where she is currently an Associate Professor of Electrical
and Computer Engineering.

Dr. Erkip received the 2004 Communications Society Stephen O. Rice
Paper Prize in the Field of Communications Theory and the NSF CAREER
award in 2001. She is an Associate Editor of IEEE Transactions on
Communications,  a Publications Editor of IEEE Transactions on
Information Theory  and a Guest Editor of IEEE Signal Processing
Magazine, Special Issue on Signal Processing for Multiterminal
Communication Systems.  She is the Technical Area Chair for the
``MIMO Communications and Signal Processing'' track of 41st Annual
Asilomar Conference on Signals, Systems, and Computers, and the
Technical Program Co-Chair of 2006 Communication Theory Workshop.
Her research interests are in wireless communications, information
theory and communication theory.
\end{biographynophoto}

\end{document}